\documentclass[iop,numberedappendix,twocolappendix]{emulateapj}
\usepackage[T1]{fontenc}
\usepackage{newtxtext, newtxmath}
\usepackage{xspace}
\usepackage{savesym}
\savesymbol{tablenum}
\usepackage{siunitx}
\restoresymbol{SIX}{tablenum}

\usepackage{url}
\usepackage{natbib}
\usepackage{graphicx}
\usepackage{color}
\usepackage{rotating}
\usepackage[breaklinks,colorlinks,citecolor=blue,linkcolor=magenta]{hyperref} 

\usepackage[all]{hypcap} %Links go to figures; breaks on deluxetables (use \capstartfalse \capstarttrue to fix it)

\newcommand{\Planck}{\textit{Planck}\xspace}
\newcommand{\Herschel}{\textit{Herschel}\xspace}
\renewcommand{\micron}{\si{\um}\xspace} % redefine \micron to use siunitx, which provides upright mu

\shorttitle{Carina Nebula Polarization Spectrum}
\shortauthors{Shariff et al.}

\begin{document}

\title{Submillimeter Polarization Spectrum of the Carina Nebula}

\author{Jamil~A.~Shariff\altaffilmark{1},
Peter~A.~R.~Ade\altaffilmark{2},
Francesco~E.~Angil\`e\altaffilmark{3},
Peter~Ashton\altaffilmark{4},
Steven~J.~Benton\altaffilmark{5},
Mark~J.~Devlin\altaffilmark{3},
Bradley~Dober\altaffilmark{6},
Laura~M.~Fissel\altaffilmark{7},
Yasuo~Fukui\altaffilmark{8},
Nicholas~Galitzki\altaffilmark{9},
Natalie~N.~Gandilo\altaffilmark{10,11},
Jeffrey~Klein\altaffilmark{3},
Andrei~L.~Korotkov\altaffilmark{12},
Zhi-Yun~Li\altaffilmark{13},
Peter~G.~Martin\altaffilmark{1},
Tristan~G.~Matthews\altaffilmark{14},
Lorenzo~Moncelsi\altaffilmark{15},
Fumitaka~Nakamura\altaffilmark{16},
Calvin~B.~Netterfield\altaffilmark{17,18},
Giles~Novak\altaffilmark{14},
Enzo~Pascale\altaffilmark{19},
Fr{\'e}d{\'e}rick~Poidevin\altaffilmark{20,21},
Fabio~P.~Santos\altaffilmark{22},
Giorgio~Savini\altaffilmark{23},
Douglas~Scott\altaffilmark{24},
Juan~Diego~Soler\altaffilmark{22},
Nicholas~E.~Thomas\altaffilmark{11},
Carole~E.~Tucker\altaffilmark{2},
Gregory~S.~Tucker\altaffilmark{12},
and Derek~Ward-Thompson\altaffilmark{25}}

\affil{} %dummy
\affil{$^1$Canadian Institute for Theoretical Astrophysics (CITA),
University of Toronto, 60 St.~George St, Toronto, ON, M5S~3H8, Canada;
\href{mailto:jshariff@cita.utoronto.ca}{jshariff@cita.utoronto.ca}}
\affil{$^2$School of Physics \& Astronomy, Cardiff University, Queens Buildings, The Parade, Cardiff, CF24~3AA, U.K.}
\affil{$^3$Department of Physics \& Astronomy, University of Pennsylvania, 209 South 33$^{\textnormal{rd}}$ St, Philadelphia, PA, 19104, U.S.A.}
\affil{$^4$Department of Physics, University of California, Berkeley,
366 LeConte Hall, Berkeley, CA, 94720, U.S.A.}
\affil{$^5$Department of Physics, Princeton University, Jadwin Hall, Princeton, NJ, 08544, U.S.A.}
\affil{$^6$National Institute for Standards and Technology, 325
Broadway, Boulder, CO, 80305, U.S.A.}
\affil{$^7$National Radio Astronomy Observatory, 520 Edgemont Rd, Charlottesville, VA, 22903, U.S.A.}
\affil{$^8$Department of Physics and Astrophysics, Nagoya University, Nagoya, 464-8602, Japan}
\affil{$^{9}$Center for Astrophysics and Space Sciences, University
of California, San Diego, 9500 Gilman Drive, \#0424, La Jolla, CA, 92093,
U.S.A}
\affil{$^{10}$Henry A.~Rowland Department of Physics \& Astronomy, Johns Hopkins
University, 3701 San Martin Drive, Baltimore, MD, 21218, U.S.A}
\affil{$^{11}$NASA Goddard Space Flight Center, 8800 Greenbelt Rd,
Greenbelt, MD, 20771, U.S.A.}
\affil{$^{12}$Department of Physics, Brown University, 182 Hope St, Providence, RI, 02912, U.S.A.}
\affil{$^{13}$Department of Astronomy, University of Virginia, 530 McCormick Rd, Charlottesville, VA, 22904, U.S.A.}
\affil{$^{14}$Center for Interdisciplinary Exploration and Research in
Astrophysics (CIERA), and  Department of Physics \& Astronomy, 
Northwestern University, 2145 Sheridan Rd, Evanston, IL, 60208, U.S.A.}
\affil{$^{15}$Division of Physics, Mathematics and Astronomy, California Institute of Technology, 1200 E.~California Blvd, Pasadena, CA, 91125, U.S.A.}
\affil{$^{16}$National Astronomical Observatory, Mitaka, Tokyo, 181-8588, Japan}
\affil{$^{17}$Department of Physics, University of Toronto, 60 St.~George St, Toronto, ON, M5S~1A7, Canada}
\affil{$^{18}$Department of Astronomy \& Astrophysics, University of Toronto, 50 St.~George St, Toronto, ON, M5S~3H4, Canada}
\affil{$^{19}$Department of Physics, La Sapienza Universit{\`a} di Roma,
Piazzale Aldo Moro 2, 00185, Roma, Italy}
\affil{$^{20}$Instituto de Astrof{\'i}sica de Canarias, E-38205, La Laguna, Tenerife, Spain}
\affil{$^{21}$Departamento de Astrof{\'i}sica, Universidad de La Laguna, E-38206, La Laguna, Tenerife, Spain}
\affil{$^{22}$Max-Planck-Institute for Astronomy, K{\"o}nigstuhl 17,
69117, Heidelberg, Germany}
\affil{$^{23}$Department of Physics \& Astronomy, University College London, Gower St, London, WC1E~6BT, U.K.}
\affil{$^{24}$Department of Physics \& Astronomy, University of British Columbia, 6224 Agricultural Rd, Vancouver, BC, V6T~1Z1, Canada}
\affil{$^{25}$Jeremiah Horrocks Institute,
University of Central Lancashire, PR1~2HE, U.K.}

\begin{abstract}
    Linear polarization maps of the Carina Nebula were obtained at 250, 350,
    and 500~\micron during the 2012 flight of the BLASTPol balloon-borne 
    telescope. These measurements are combined with \Planck 850~\micron data
    in order to produce a submillimeter spectrum of the polarization fraction of
    the dust emission, averaged over the cloud. This spectrum is flat to within
    $\pm$15\% (relative to the 350~\micron polarization fraction). In particular, there is no evidence for a pronounced minimum of the spectrum near
    350~\micron, as suggested by previous ground-based measurements of other
    molecular clouds. This result of a flat polarization spectrum in Carina is
    consistent with recently-published BLASTPol measurements of the
    Vela C molecular cloud, and also agrees with a published model for an
    externally-illuminated, dense molecular cloud by Bethell and
    collaborators. The shape of the spectrum in Carina does
    not show any dependence on the radiative environment of the dust, 
    as quantified by the \Planck-derived dust temperature or dust optical 
    depth at 353 GHz.
\end{abstract}

\maketitle
%\tableofcontents
\section{Introduction}
\label{sec:intro}
\normalsize

Dust grains in the interstellar medium (ISM) have long been known to linearly
polarize background starlight at visible and near-infrared
wavelengths~\citep{hall49, hiltner49}. The thermal emission from these dust
grains in the far-infrared and submillimeter portion of the spectrum is
observed to be linearly polarized as well~\citep{cudlip82, hildebrand84}. 
It is believed that the
aspherical, spinning dust grains align with their long axes perpendicular
to the direction of the local ISM magnetic field. The mechanism for the alignment of the dust grains is
an area of active research, but recently the theory of radiative alignment
torques (RATs) has gained some observational support. Under the RAT
mechanism~\citep{lazarian07}, an external radiation field is able to provide a net torque to
spin up dust grains, which develop a net magnetic dipole moment that
aligns with the direction of the local magnetic field. A detailed review
is given in~\citet{andersson15}. In particular, they note that RAT
alignment requires the presence of a radiation field with wavelengths less
than the grain diameter. This condition predicts a lower alignment
efficiency, and hence a lower polarization fraction, with increasing dust
extinction.

The aligned dust grains
preferentially absorb and emit light that is polarized in the direction parallel
to their long axes. As a result, transmitted starlight is
preferentially polarized in the direction parallel to the local magnetic
field, whereas the dust thermal emission is polarized in the direction
perpendicular to the magnetic field. Dust polarimetry therefore allows the
directions of the plane-of-sky projection of the ISM magnetic field to be
traced. This measurement can serve as a powerful tool for investigating
the role played by magnetic fields in the early stages of star formation. 
This is true particularly when the polarization is
measured in emission at longer wavelengths, where it can be traced into the
interiors of the dense clouds of molecular gas where star formation takes
place. Measurements of polarized dust emission in the diffuse ISM are also
important, since this emission is a source of foreground contamination to
studies of the polarization of the Cosmic Microwave Background (CMB)
radiation. Data that provide a better understanding of the 
variation of the dust polarization fraction with wavelength, and with dust
environment are important to both of these scientific applications.  

The submillimeter \emph{polarization spectrum} $p(\lambda)$ is the
linear polarization fraction of the thermal dust
emission as a function of wavelength (polarization fraction being defined in Section~\ref{sec:polarimetry}). Typically, the spectrum is divided
by the polarization fraction at a reference wavelength $\lambda_0$.
This normalization removes
the dependence on the inclination angle of the magnetic field relative to
the line of sight, and on any other unknown factors that would affect the
observed polarization fraction across all bands. The relevant observable
for the study of dust grain alignment is then the \emph{shape} of this
normalized spectrum $p(\lambda)/p(\lambda_0)$.

A number of models have
been developed attempting to predict the shape of the polarization spectrum
over the submillimeter spectrum, in various column density regimes.
\citet{draine09} investigated the polarization of dust in the diffuse ISM
using models of aspherical silicate and graphite grains that were
constrained to reproduce the observed dust extinction and polarization of
starlight as a function of wavelength. They produced model polarization spectra
that were rising from 100~\micron to 1000~\micron. For a scenario in which
the dust is diffuse enough that all grains are exposed to the same
interstellar radiation field, it is not unreasonable to assume that the dust
temperature will depend only on grain material and size, and not on
physical location. Under this assumption, the rising spectrum can be
explained, at least in part, in terms of an anti-correlation
between dust temperature and grain alignment. Larger dust grains are known
empirically to be better-aligned~\citep{kim95}. However, larger dust
grains also have higher emissivity and thus tend towards lower equilibrium
temperatures. Therefore, it is the well-aligned lower-temperature grains, whose emission
peaks at longer wavelengths, that contribute predominantly to the
polarization fraction in this scenario. 

In contrast to this diffuse ISM study, \citet{bethell07} modelled polarized dust emission in
\emph{dense}, clumpy molecular clouds and cores using simulations of
magnetohydrodynamic turbulence. They found a polarization fraction that is
largely flat over submillimeter wavelengths, and only begins falling off towards
the far-infrared, decreasing by a factor of two in percentage polarization
between 250~\micron and 100~\micron. A possible explanation for this lack
of variation is known as the extinction-temperature-alignment correlation
(ETAC) effect. Under this effect, grains in the interior of dense
molecular clouds are more shielded from the interstellar radiation field
than grains on the surface of these clouds. If the RAT mechanism is
correct, the shielded interior grains should be both colder and less
well-aligned, while the surface grains should be warmer and
better-aligned. This is the inverse temperature-alignment correlation from
the one discussed above for the diffuse ISM. The net result is that the
average temperature of the grains contributing to polarized emission is
closer to the average temperature of all grains, leading to a flatter
polarization spectrum. Additional discussion of the ETAC effect appears
later in this section, and in Section~\ref{sec:disc}.

Previous ground-based observations of the submillimeter polarization
spectrum in dense clouds and cores have found large ratios in the
polarization fraction between different bands. These observations
typically span $ < 0.01~\si{\deg}^2$, around bright sources. As shown
in~\citet{vaill12}, the combination of ground-based
measurements of different targets in different bands produces a V-shaped
polarization spectrum that falls very steeply in the far-infrared, shows a
pronounced minimum at 350~\micron, and rises very steeply towards
millimeter wavelengths (see Figure~\ref{fig:money}). The models described above are not able to account
for the steepness of the observed slopes, nor the overall magnitude of the
variation.

More recently, high-sensitivity, wide-area mapping observations by the
Balloon-borne Large Aperture Submillimeter Telescope for Polarimetry 
(BLASTPol) at 250, 350, and 500~\micron have produced
polarization measurements for molecular-cloud targets in the
Galactic plane. For example, \citet{gand16} present the submillimeter polarization
spectrum of the Vela C giant molecular cloud (GMC). \citet{ashton18}
compute the first submillimeter polarization spectrum of a translucent
molecular cloud near to Vela C on the sky, but having approximately an
order of magnitude lower column density. Both of these studies combined
data from the three BLASTPol bands with data from the \Planck High
Frequency Instrument (HFI) 353 GHz (850~\micron) band. Both analyses resulted
in polarization spectra that were flat to within $\pm$15\% in
$p(\lambda)/p(\lambda_0)$ over these four
bands.

It should be noted that~\citet{ashton18} attempted to model the
ETAC effect analytically using two different populations of grains---bulk
and surface---having different temperature-size distributions and alignment
fractions. They found that in their observed range of column densities, 
this implementation of the ETAC effect was not strong enough to account
for the observed flatness of the polarization spectrum of the translucent
molecular cloud. In other words, the diffuse ISM models
of~\citet{draine09} with no shielding should be applicable to the cloud observed
by~\citet{ashton18}. However, these models are rising with wavelength, rather
than flat, and can disagree with the translucent cloud data by up to 
30\% at 250~\micron. Further modelling work, such as that
of~\citet{guillet18} attempts to produce flatter polarization spectra by
varying the composition, porosity, and oblateness of the aligned grains.

To summarize some of the most recent developments in polarization spectrum
analysis: combined BLASTPol/\Planck results have produced flat
polarization spectra in two different clouds of very different densities.
These results disfavor some of the~\citet{draine09} models in the diffuse 
case, and are in sharp
contrast with previous ground-based measurements in the case of dense
clouds. Thus, it has become even more important to study the
polarization spectra of other targets, preferably having a range of
different cloud environments, so as to further explore these
discrepancies, and to better inform grain-alignment models.

The Carina Nebula (NGC 3372), the largest and highest-surface-brightness
nebula in the southern sky, appears in visible light as a giant
\textsc{H\,ii} region spanning several square degrees. Located at an
estimated distance of 2.3 kpc~\citep{allen93, smith06}, the nebula and
surrounding molecular cloud are part of a GMC
complex spanning some 150~pc. %In the past two decades, thanks to multiwavelength observations, a picture of Carina has emerged in which active star formation is occurring at the periphery of the nebula, triggered by feedback in the form of intense ultraviolet radiation and stellar winds from dozens of massive O-type stars located in the nebula's central clusters. At the same time this feedback can disrupt the molecular cloud from which the stars formed, removing the raw material necessary for further star formation. A comprehenseive review is given by~\citet{smith08} who note that Carina serves as an ideal laboratory in which to study the effects of feedback from massive stars. 
In the context of the BLASTPol observations, the Carina GMC is perhaps the
most active and evolved source observed, the other targets being
relatively more quiescent molecular clouds. An overview of the 
the Carina molecular cloud
complex, including the structure of the submillimeter emission, 
is given in~\citet{li2006}. As they note, the central open
clusters, Trumpler 14 and 16, contain an unusual concentration of massive
stars, including $\eta$~Carinae, and six of the 17 O3-type stars in
the Galaxy that were known at that time. In contrast, the most massive
sources of excitation for the 
\textsc{H\,ii} region RCW~36 in Vela C have been measured to be two stars
of type O9 and O9.5~\citep{ellerbroek13}. For this reason, comparisons of
submillimeter polarimetric observations of Carina with other molecular
clouds, such as Vela C, might be regarded as a way to probe the effects of radiative environment and internal heating on dust grain alignment, particularly in the context of the RAT mechanism. 

In this paper, BLASTPol polarization data from the Carina Nebula at 250,
350, and 500~\micron are presented along with \Planck HFI 353 GHz
(850~\micron) data from the same region. A submillimeter polarization
spectrum of Carina is produced over these bands following a similar, but
independent analysis to that of~\citet{gand16} for Vela C. Section~\ref{sec:obs} describes the BLASTPol instrument, the 2012
science flight, and the steps of the
data analysis, including a detailed description of the polarimetric analysis.
Section~\ref{sec:results} contains the main results of the polarization
spectrum analysis for Carina. The implications of these results are
discussed in Section~\ref{sec:disc}, and the overall findings of this paper are
summarized in Section~\ref{sec:summ}.

\section{Observations and Data Analysis}
\label{sec:obs}
\subsection{The BLASTPol Instrument \& 2012 Flight}
The BLASTPol instrument was a stratospheric balloon-borne polarimeter. The
telescope consisted of a 1.8~m aluminum primary mirror and a 40~cm aluminum
secondary mirror. Light from the secondary passed into
a re-imaging optics box that was cooled to approximately 1.5~K within a cryostat employing liquid
nitrogen and liquid helium cooling stages. In the optics box, an achromatic
half-wave plate allowed the linear polarization of the incident
radiation to be rotated periodically. Dichroic beam splitters then directed the radiation onto one of three
focal planes consisting of 300~mK feedhorn-coupled bolometric detectors 
operating at 250, 350, and 500~\micron. These focal plane arrays were very
similar to those that were flown on the \Herschel SPIRE
instrument~\citep{griffin03}, but with the addition of lithographed
polarizing grids placed in front of each feedhorn array. More details of the
BLASTPol instrument can be found in~\citet{tyr14}. 

BLASTPol was launched from the vicinity of McMurdo Station, Antarctica on
2012 December 26. It conducted observations for 12.5 days at a mean
altitude of 38.5~km above sea level. The duration of observations was
limited by the boil-off time of the liquid helium. 

The highest BLASTPol observation time for a single target was devoted to 
the Vela C GMC. However, observations of the Carina Nebula
also took place, totalling 4.2 hours, and covering approximately
$2.5~\si{\deg}^2$. 
The coverage area was chosen to overlap with the observation region and
three-point-chop reference regions of the ground-based Submillimeter
Polarimeter for Antarctic Remote Observations (SPARO).
SPARO observed Carina and several other GMCs at 450~\micron~\citep{li2006}.

The raw output from the experiment consisted of streams of time-ordered
data (TOD), one per bolometer. A number of pre-processing steps had to be
applied to the TOD before they could be binned into maps of the sky.
This time-domain pre-processing, along with modelling of the 
in-flight beam, and the estimation of instrumental
polarization, are described in~\citet{fissel16}.

\subsection{Map Making}

\begin{figure*}[htbp!]
%\begin{figure*}[p]
%\epsscale{1.0}
\plotone{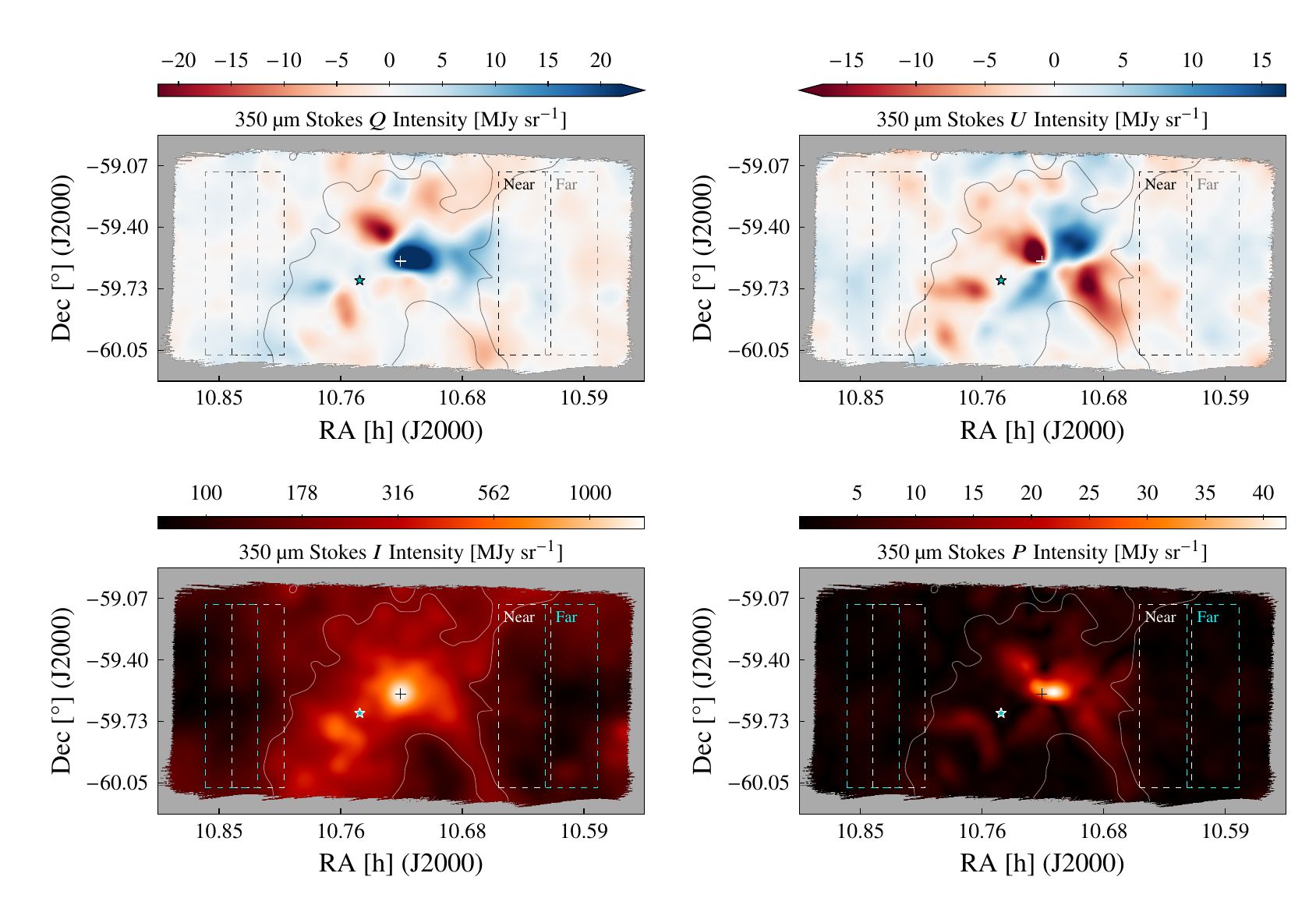}
\caption{BLASTPol 350~\micron intensity maps of the Carina Nebula in
    Stokes $Q$ (top left), $U$ (top right), $I$ (bottom left), and 
    $P$ (bottom right). The
    dashed rectangles show the Near (white or black) and Far (cyan or gray) 
    reference regions. The data used in this analysis are enclosed by the 
    solid contour, which is determined by a threshold cut on Stokes $I$
    at 850~\micron (see text). The crosshair shows the location of the peak
    in $I$, while the star-shaped marker shows the position of $\eta$~Car. 
    Note that the $I$ map is displayed on a
    logarithmic scale, while the other maps are on a linear scale.\label{Imap}}
\end{figure*}

BLASTPol maps of the $I$, $Q$, and $U$ Stokes parameters
(Figure~\ref{Imap}) were produced 
using TOAST (Time-Ordered
Astrophysics Scalable Tools)\footnote{\url{https://github.com/tskisner/TOAST}},
a set of code for map making and simulation that can be used serially or
with OpenMP/MPI parallelization.
The TOAST generalized least-squares (GLS) solver was used, which
iteratively inverts the map maker equation using the preconditioned
conjugate gradient method. The map maker's input noise model came from
per-bolometer TOD power spectral densities estimated from data obtained
while observing a low-signal region of sky. It was not necessary for the
noise model to 
include non-stationarity or detector-to-detector noise correlations. Using
the input noise model, the map maker produces a $3\times3$ matrix of the $(I, Q, U)$ covariances for each pixel. 
The TOAST maps were produced using a $10^{\prime\prime}$ pixelization.
Data from input \Planck HFI 850~\micron
all-sky maps\footnote{The input \Planck maps at 353~GHz were obtained for the
second data release (PR2 2015) from the Planck Legacy Archive:
\url{http://pla.esac.esa.int}.}, were processed using coordinate
information from TOAST to produce 850~\micron
maps of the Carina Nebula with the same pixelization, angular extent, and map 
projection as the BLASTPol maps. All of the TOAST Carina maps were produced in Equatorial
coordinates, which is therefore the coordinate system to which $Q$ and $U$
are referenced throughout this work (Section~\ref{sec:polarimetry}). 
For this analysis, the BLASTPol signal maps were
smoothed to a beam size of $4^\prime.8$ full width at half-maximum (FWHM), 
to match the resolution of the \Planck data. This is also well above
the scale  of irregularities that were observed in
the BLASTPol beam shape~\citep{fissel16}. Lucy-Richardson iterative
deconvolution~\citep{lucy74, richardson72} was used to deconvolve a model of the BLASTPol beam from a
symmetric $4^\prime.8$ FWHM Gaussian beam. The result of this
deconvolution was then used as the smoothing kernel with which the BLASTPol 
signal maps were convolved. The covariance maps were smoothed with
the square of this normalized kernel.

\subsection{Calibration}
\label{sec:cal}

The TOAST BLASTPol maps are initially produced in the same units as the
    TOD: raw ADC counts. They must be calibrated into physical units of
    MJy\,sr$^{-1}$. Calibration is achieved using a dust spectral energy
    distribution (SED) obtained from the \Planck all-sky dust model
    described in~\citet{planck_xi14}. This model is also discussed in more
    detail in Section~\ref{sec:env}. The model dust morphology is defined
    by the optical depth at a reference frequency. Using the model SED,
    the dust total intensity at this frequency is scaled to and integrated
    over top-hat approximations to each of the three BLASTPol bands. This procedure produces model maps for the dust Stokes $I_\lambda$ at $\lambda$ =  250, 350, and 500~\micron. These model maps are fitted to the actual BLASTPol maps in these three bands according to
\begin{equation}
\label{eq:cal}
I_\lambda(\mathrm{BLASTPol}) = A_\lambda \left(I_\lambda(\mathrm{model}) + B_\lambda\right).
\end{equation}

The BLASTPol maps $Q_\lambda$ and $U_\lambda$ were calibrated using the
    above gain as well ($A_\lambda^{-1}$ in MJy\,sr$^{-1}$\,count$^{-1}$).
    However, they also had to be divided by $\gamma_\lambda$, the measured
    instrumental polarization efficiencies in each waveband, which are
    reported in~\citet{tyr14}. While this procedure corrected the map
    slopes, the polarization maps also had arbitrary DC offsets
    $B_\lambda$ that could not be determined using Equation~\ref{eq:cal},
    because equivalent dust model maps $Q_\lambda(\mathrm{model})$ and
    $U_\lambda(\mathrm{model})$ do not exist. As discussed in the next
    section, polarization spectrum analysis normally proceeds by
    subtracting diffuse background emission from the maps before computing
    polarization quantities. However, a case with no background
    subtraction is presented here as well, for which the determination of
    the DC offsets is important. To determine the offsets, the \Planck HFI
    850~\micron map\footnote{It should be noted that, by default, the submillimeter emission in the \Planck 850~\micron maps is expressed in units of temperature deviation in kelvins from the 2.725~K CMB blackbody (K$_\mathrm{CMB}$). For this analysis, the HFI maps were converted to MJy\,sr$^{-1}$ using the color-corrected conversion constant of 246.543~MJy\,sr$^{-1}$\,K$_{\mathrm{CMB}}^{-1}$ from Table 6 of~\citet{planck_ix14}.} of dust emission was color-corrected by scaling it to the dust model map in each of the BLASTPol bands:
\begin{equation}
\label{eq:cal2}
I_\lambda(\mathrm{model}) = a_\lambda \left(I_{850}(\mathrm{HFI}) + b_\lambda\right).
\end{equation} In principle, the linearity of this scaling requires that
    the dust be isothermal at temperature $T_d$, and that it have an
    emissivity with consant power-law spectral index $\beta_d$
    (Section~\ref{sec:env}). In practice, it was found that rejection of
    map pixels that were high-temperature outliers in the all-sky dust
    model was sufficient for good linearity. Model maps
    $Q_\lambda(\mathrm{model})$ and $U_\lambda(\mathrm{model})$ could then be obtained by color-correcting $Q_{850}$ and $U_{850}$ to the BLASTPol bands using scale factor $a_\lambda$. The DC offsets of the calibrated BLASTPol polarization maps were then obtained by plotting pixel values of $Q_\lambda(\mathrm{BLASTPol})$ versus $Q_\lambda(\mathrm{model})$, and likewise for $U_\lambda$.

\subsection{Diffuse Emission Subtraction}

Care must be taken to ensure that the polarized intensity observed at each
sightline in this analysis is restricted to emission from the molecular cloud
itself, and does not include a component from diffuse Galactic dust in the
foreground or background. Contamination from diffuse Galactic emission
could have a significant effect on the measured polarization, since
diffuse emission has been shown, on average, to have a higher linear
polarization fraction than emission from denser regions within molecular
clouds~\citep{planck_int_xix15}.

Typically, to correct for diffuse emission, a reference region adjacent to the cloud (but
outside of it) is chosen. In each of $I$, $Q$, and $U$, the mean intensity 
within the reference region
is assumed to be a level of diffuse emission from foreground or
background dust that applies uniformly to the sightlines within the cloud
as well. This intensity level is subtracted from its respective Stokes map
before proceeding with the polarization analysis. For this analysis of
Carina, two narrow vertical rectangular reference regions were chosen that
bracket the cloud on its east and west sides. The intensity level subtracted
was the mean intensity of all the pixels lying within the two rectangular
regions. As shown in Figure~\ref{Imap}, two different sets of two rectangles were
selected, a closer pair  called the ``Near'' reference region, and a pair
farther out in RA, referred to as the ``Far'' reference region. All the
analysis presented herein was repeated for both the Near and Far reference
regions, in order to evaluate the dependence of the result on the choice of
reference region. Furthermore, for the purpose of comparison, we present an additional analysis for which no diffuse emission subtraction is carried out. 

A large-scale, roughly north-south systematic gradient exists in the BLASTPol 
Carina Nebula maps due to receiver $1/f$ noise. This gradient is corrected using
\Herschel SPIRE $I$ maps of the same region in the same bands
(see Appendix~\ref{sec:herschel}). However, choosing the reference regions to be 
elongated vertically, and averaging over them, mitigates the effects
of any residual gradient still present after this correction.

In selecting reference regions that lie just outside (or well outside) the
molecular cloud, the question arises of which map pixels are associated
with the cloud in the first place. These pixels were chosen by applying a
threshold cut on Stokes $I$ at 850~\micron (the intensity in this waveband
being used as a proxy for dust column density). The cloud is then defined as the 
region enclosed by the contour where $I_{850} = (3/2)\langle I_{850}
\rangle_{\mathrm{Far}}$, with the angle brackets denoting the mean
intensity over the Far reference region. This contour is overlaid on the
maps in Figure~\ref{Imap}; pixels outside of it were excluded from the
polarization spectrum analysis. The ratio of 3/2 was chosen because it
is comparable to the 850~\micron intensity ratio between the Vela C cloud 
regions defined by~\citet{hill2011}, and the reference regions
used in the BLASTPol polarization spectrum analysis of Vela
C~\citep{gand16}\footnote{The cloud-to-reference-region intensity ratio of 3/2 chosen for Carina corresponds the closest to Vela C when using the ``aggressive'' or ``intermediate'' reference regions for Vela C that are defined in~\citet{gand16}.}. Data that lie outside vertical
lines coinciding with the inner
vertical edges of the reference region rectangles on either side of the map have
also been excluded from the analysis in both the Near and Far cases. In the case with no background subtraction, the Far reference region rectangles are used for this purpose.

After carrying out the diffuse emission subtraction (if any) the Stokes parameter maps are
spatially-downsampled by a factor of 15 using constant-value interpolation. This
step increases the pixel size from $10^{\prime\prime}$ to $2^\prime.5$, 
roughly Nyquist-sampling the $4^\prime.8$ beam of the smoothed maps. The pixel covariance maps are downsampled in the same manner.
\begin{figure}[htbp!]
%\begin{figure}[p]
\epsscale{1.0}
\plotone{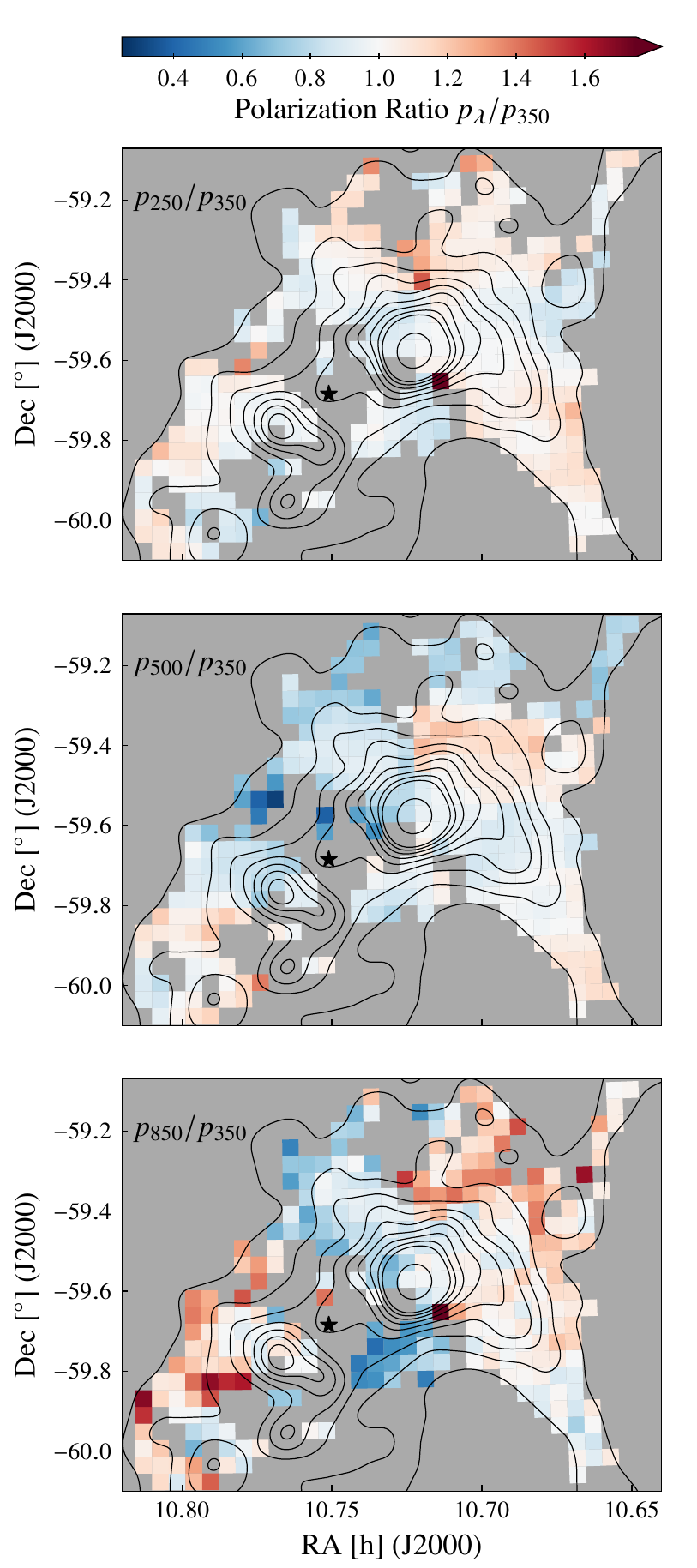}
    \caption{Spatial maps of the polarization ratios $p_{250}/p_{350}$
    (top), $p_{500}/p_{350}$ (middle), and $p_{850}/p_{350}$ (bottom). The
    contours show percentages of the 350~\micron peak intensity ranging 
    from 5\% to 50\% in 5\% increments, along with a 75\% contour. The
    black star-shaped marker shows the location of $\eta$~Car.
        \label{fig:pol_ratio_maps}}
\end{figure}

\begin{figure*}[tbp!]
%\begin{figure*}[p]
\epsscale{1.0}
\plotone{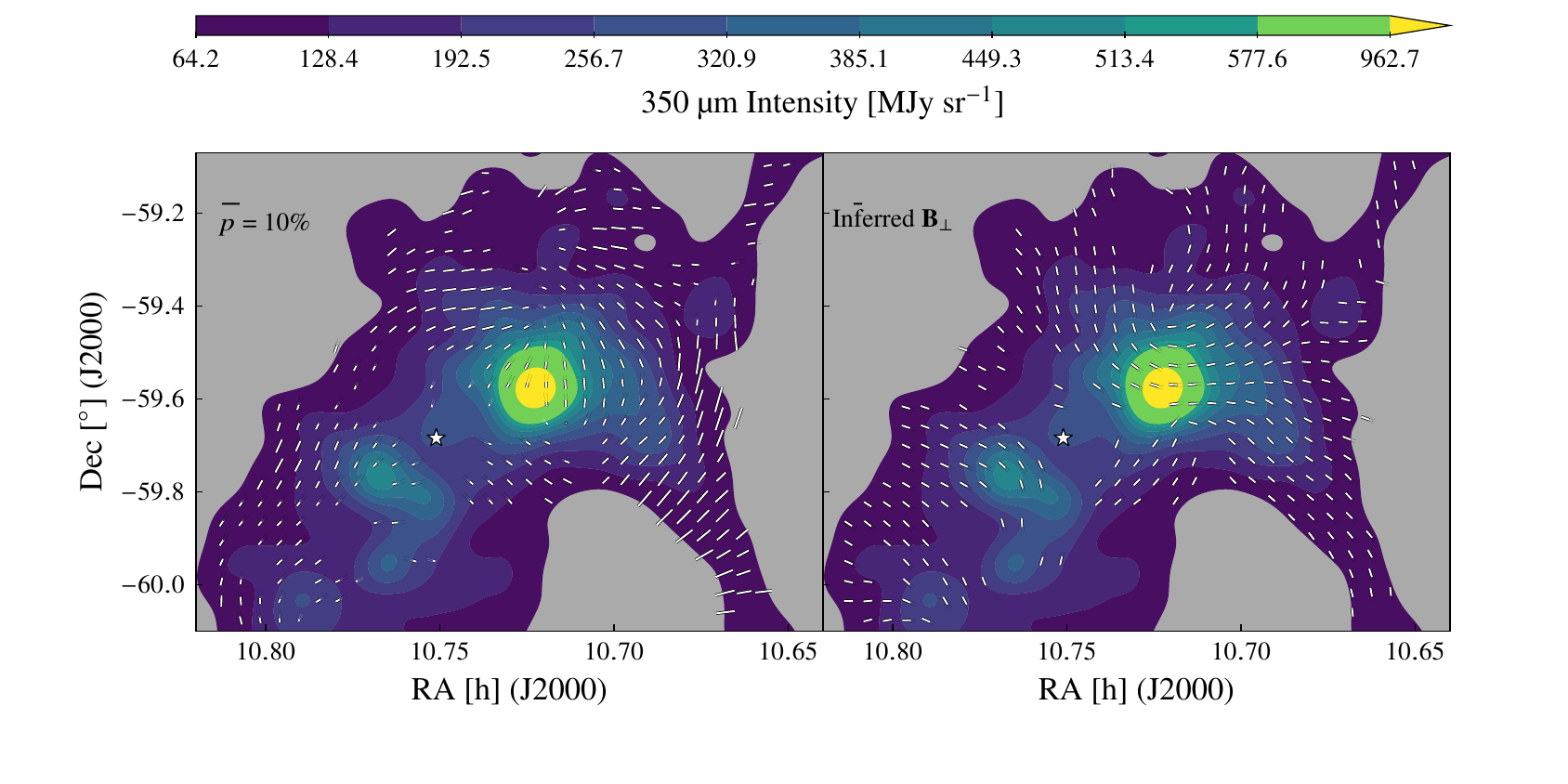}
    \caption{Left panel: polarization pseudo-vectors showing the 
    direction $\psi$ of the linearly-polarized
    radiation measured by BLASTPol at 350~\micron in Carina, 
    for those sightlines surviving the data cuts
    (see text). The lengths of the pseudo-vectors are scaled to show the
    fractional linear polarization $p$ for each sightline, 
    as a percentage of the total intensity. The length scale is
    given by the key in the upper left. 
    Right panel: pseudo-vectors showing the corresponding
    directions of the projected magnetic field $\mathbf{B}_\perp$ within
    the cloud. In this case,
    the pseudo-vectors are all the same length, in order to show the large-scale
    structure of the field more clearly. The color scale is for the filled
    contours in both panels, which show the 350~\micron intensity $I$ in
    MJy\,sr$^{-1}$. The star-shaped markers indicate $\eta$~Car. \label{vectors}}
\end{figure*}

\subsection{Polarimetry}
\label{sec:polarimetry}

The Stokes parameters are used to compute the net linear polarization of the dust
emission
\begin{equation}
    \label{eq:P}
    P = \sqrt{Q^2 + U^2},
\end{equation}
as well as the fractional linear polarization
\begin{equation}
    \label{eq:pfrac}
    p = \frac{P}{I}.
\end{equation}
The angle $\psi$ defining the direction of the linear polarization on the
sky is given by
\begin{equation}
    \label{eq:psi}
    \psi = \frac{1}{2}\arctan(U,Q).
\end{equation}
where the two-argument form of the arctan function is used in order to
evaluate the angle quadrant properly. The IAU polarization angle
convention is used; for a polarization pseudo-vector viewed
on the sky in Equatorial coordinates, $\psi$
increases counter-clockwise from $0\si{\degree}$ in the the north-south direction, and ranges
from $-90\si{\degree}$ to $+90\si{\degree}$.

The TOAST pixel covariances are used to compute the variances in these
quantities, $\sigma_p^2$ and $\sigma_\psi^2$, using error propagation
(see Appendix~\ref{sec:err}). Since $P$ and $p$ are restricted to positive
values, any noise in $Q$ and $U$ positively biases these quantities. The
polarization fraction is de-biased approximately using a rudimentary method that is
acceptable for high signal-to-noise in $p$~\citep{wardle74, montier15}:
\begin{equation}
    p_{\mathrm{db}} = \sqrt{p^2 - \sigma_p^2}.
\end{equation}
All of the $p$ values used in the polarization spectrum analysis
(Section~\ref{sec:results}) have been de-biased in this way. Note, however, that
the map of $P$ shown in the lower right panel of Figure~\ref{Imap} is
presented for visualization purposes only, and has not been de-biased.

Once maps of $p$ and $\psi$ have been produced, data cuts are applied
that have been customary for submillimeter polarization spectrum analyses
in the literature. 
The first is a signal-to-noise cut on map pixels using a threshold of 
$p_{\mathrm{db}} > 3\sigma_p$. Only pixels for which this condition holds simultaneously in all four bands are kept. The second data cut is intended
to mitigate the circumstance in which the different wavebands sample
different cloud components along the line of sight, each with differing
line-of-sight components of the  magnetic field, and hence differing polarization
fractions. This situation would lead to artificial variation with
wavelength in the measured
polarization spectrum for the sightline in question:
variation which is not intrinsic to any particular physical location within
the cloud. Under the assumption that a sightline having a plane-of-sky component of
$\mathbf{B}$ that is constant with wavelength implies (at least in a
statistical sense) a constant line-of-sight component of $\mathbf{B}$ as
well, the
condition is imposed that the difference $|\Delta \psi| < 15\si{\degree}$ between
\emph{any} two of the four wavebands. The stringent maximum
angle difference of $10\si{\degree}$ used in past analyses,
including~\citet{gand16}, was relaxed for Carina, in order to include more
sightlines in the analysis.

After the downsampling and data cuts,
polarization data remained for 314, 285, and 270 sightlines respectively
for the three cases of diffuse emission subtraction using the Far
reference region, subtraction using the Near reference region, 
and no diffuse emission subtraction at all. Figure~\ref{fig:pol_ratio_maps} shows maps of the polarization
fraction \emph{ratios} for those pixels surviving the data cuts in the Far
reference region case. These are
the ratios $p_\lambda / p_{350}$, for $\lambda \in \{250, 500,
850\}$~\micron. These discrete values sample the polarization spectrum over
this wavelength range, and are analyzed in detail in the next section.

For all map pixels surviving the data cuts, the left panel of Figure~\ref{vectors} represents the 
linear polarization at 350~\micron as pseudo-vectors of length $p$ and 
direction $\psi$. The pseudo-vectors are overlaid on filled contours
showing the 350~\micron intensity over the cloud. The ``polarization-hole
effect'' is evident here, in which the polarization fraction is lower near bright intensity peaks. This effect has been noted in past submillimeter
polarimetry observations~\citep{matthews01}. The right panel of
    Figure~\ref{vectors} shows the corresponding inferred directions (but not magnitudes) 
of the plane-of-sky component of the magnetic field, $\mathbf{B}_\perp$.
These directions are rotated by $90\si{\degree}$ relative to the
$\mathbf{E}$-field polarization direction of the radiation. 

\section{Results}
\label{sec:results}
\subsection{Median Polarization Ratios}
\label{sec:polratios}
\begin{figure}[htbp!]
%\begin{figure}[p]
\epsscale{1.0}
\plotone{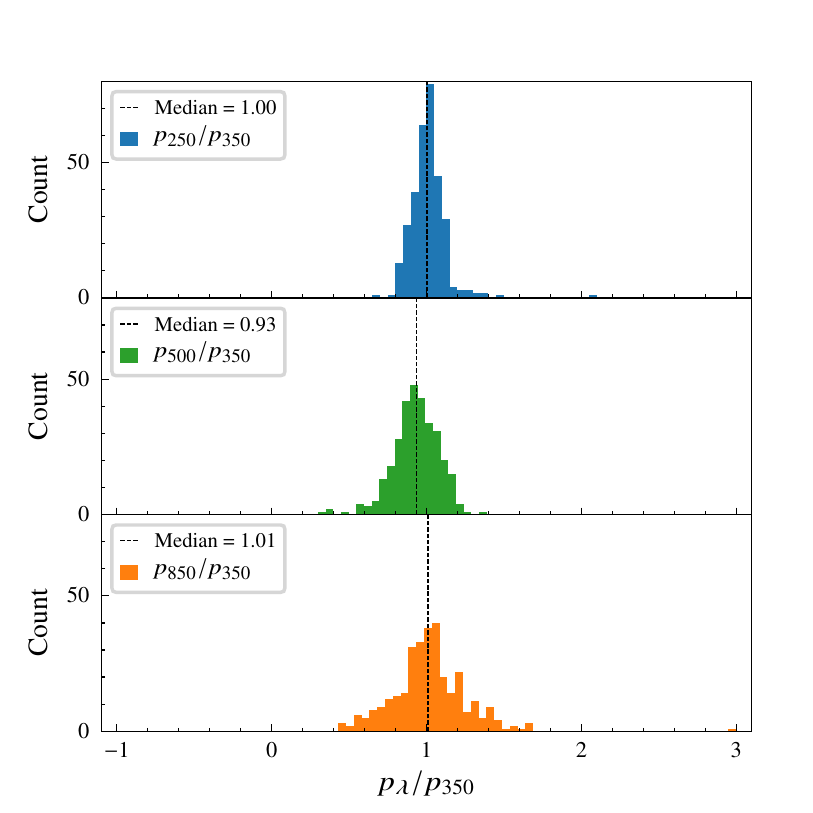}
    \caption{Histograms of the polarization ratios $p_\lambda/p_{350}$ for
    diffuse emission subtraction using the Far reference region.
    The dashed vertical lines show the median of each
    distribution.\label{fig:pol_ratio_hists}}
\end{figure}

Figure~\ref{fig:pol_ratio_hists} shows histograms of the polarization
ratios $p_\lambda/p_{350}$, specifically the distributions of these ratios over the cloud for the case of diffuse 
emission subtraction using the Far reference region. For consistency with~\citet{gand16}, the widths of the distributions were
quantified using the median absolute deviation (MAD), defined as
\begin{equation}
    \mathrm{MAD} \equiv \mathrm{median}\left(|x_i -
    \mathrm{median}(x_i)|\right),
\end{equation}
where the quantities $x_i$ are the measurements in question.
Table~\ref{tab:med_mad} lists the median ratios and MADs for all three
    types of diffuse emission subtraction. Although the two cases with
    background subtraction
show a slight minimum at 500~\micron, none of the results are significantly
different from a flat spectrum, with a polarization ratio of unity to
within 15\% in each
band. This result is independent of the method of diffuse emission
subtraction: a very similar outcome to that of~\citet{gand16}. 
\renewcommand{\arraystretch}{1.25}
%\begin{table*}[htb!]
\begin{table}[htbp!]
    \caption{Medians and MADs of Polarization Ratios
    ($p_\lambda/p_{350}$).\label{tab:med_mad}}
\begin{center}
\begin{tabular}{lccc}
\hline
\hline 
Diffuse Emission&$250~\micron$&$500~\micron$&$850~\micron$\\
Subtraction Method&&&\\
\hline
Far&$1.00\pm0.06$&$0.93\pm0.10$&$1.01\pm0.12$\\
Near&$1.02\pm0.06$&$0.93\pm0.08$&$0.99\pm0.12$\\
None&$1.14\pm0.08$&$0.96\pm0.08$&$0.95\pm0.12$\\
\hline
\end{tabular}
\end{center}
%\label{table:medians}
\end{table}

\subsection{Polarization Ratios from Scatter Plots of $p_\lambda$
vs.~$p_{350}$}
    \label{sec:scatter}

\begin{figure}[htbp!]
%\begin{figure}[p]
\epsscale{1.0}
\plotone{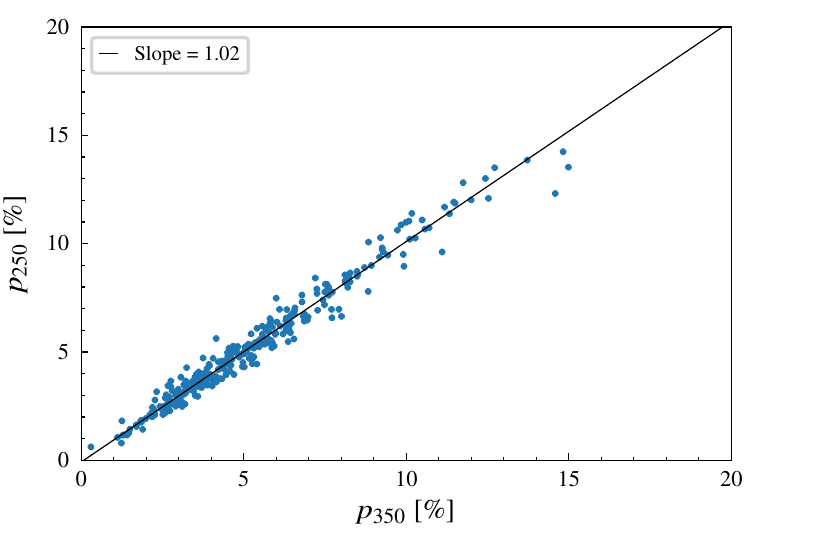}
\plotone{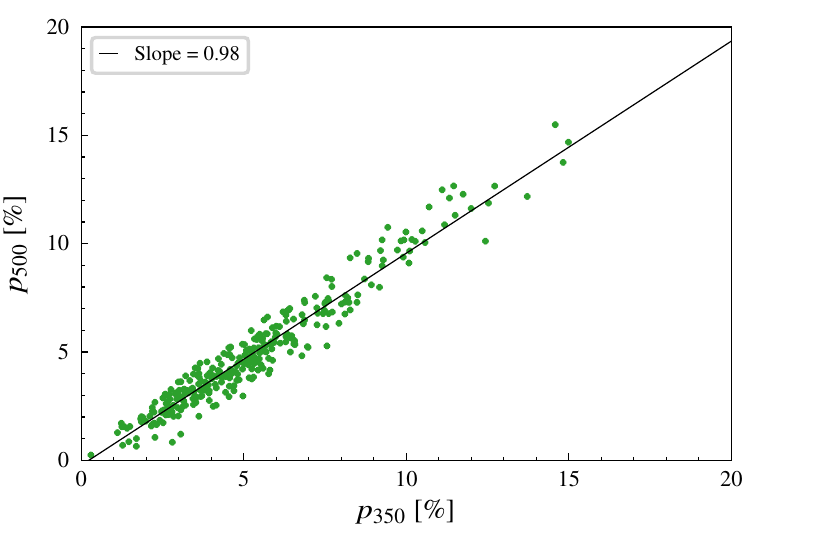}
\plotone{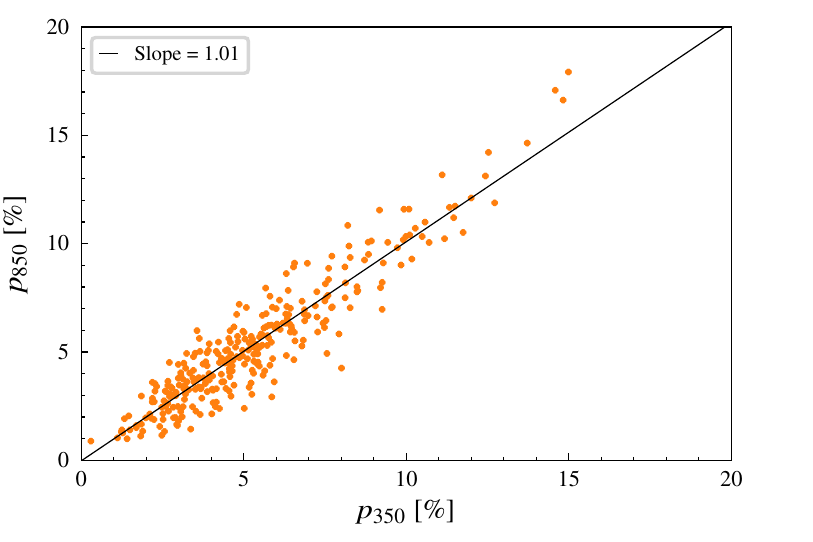}
    \caption{Linear fits to scatter plots of $p_\lambda$ versus~$p_{350}$
    for $\lambda$ = 250~\micron (top), $\lambda$ = 500~\micron (middle),
    and $\lambda$ = 850~\micron (bottom). These plots are shown for diffuse
    emission subtraction using the Far reference region.\label{fig:lin_fits}}
\end{figure}

An alternative method for determining the polarization ratios, averaged
over the cloud, is to produce linear fits to scatter plots of
$p_\lambda$ versus $p_{350}$. The polarization spectrum then consists of the best-fit 
linear slopes as a function of wavelength. For this fitting procedure, the least
absolute deviation was used to optimize the fit parameters. This
method is more robust to outliers than least-squares fitting. For each fit, an uncertainty on the slope was estimated using
bootstrap resampling~\citep[p.~691]{press1992}. The fit was repeated for each of 10,000
random selections of the data points (with replacement), and the
uncertainty was taken to be the standard deviation of this ensemble of fit
parameter values. The linear fits are shown in 
Figure~\ref{fig:lin_fits} for the case of diffuse emission subtraction
    using the Far reference region.

    Table~\ref{tab:lin_fit} lists the
    slopes of the linear fits to each waveband, along with their
    uncertainties, for the cases of background subtraction using the Far
    reference region, using the Near reference region, and for the case of
    no background subtraction. For the Near and Far cases, the feature of 
    a very slight minimum at
    500~\micron occurs using this method, just as it did for the median
    ratios. The polarization spectra obtained using the linear
    fitting are once again flat to within $\pm$15\%. 

\begin{table}[htbp!]
    \caption{Slopes of Linear Fits to
    $p_\lambda$ versus $p_{350}$.\label{tab:lin_fit}}
\begin{center}
\begin{tabular}{lccc}
\hline \hline 
Diffuse Emission&$250~\micron$&$500~\micron$&$850~\micron$\\
Subtraction Method&&&\\
\hline
Far&$1.02\pm0.01$&$0.98\pm0.02$&$1.01\pm0.02$\\
Near&$1.05\pm0.01$&$0.95\pm0.02$&$1.03\pm 0.03$\\
None&$1.14\pm0.02$&$1.02\pm0.02$&$0.90\pm0.04$\\
\hline
\end{tabular}
\end{center}
\end{table}

\subsection{Fits to $p(\lambda)$}\label{sec:lfits}
\label{sec:model}

Two different functional forms for $p(\lambda)$ were fitted to the
per-pixel $p$ measurements across the bands: a
power law
\begin{equation}
    \label{eq:pwrlaw}
    p(\lambda) = a_1 \left(\frac{\lambda}{\lambda_0}\right)^{b_1},
\end{equation}
and a second-order polynomial

\begin{equation}
    \label{eq:quadlaw}
    p(\lambda) = a_2 \left[b_2 \left(\lambda-\lambda_0\right)^2 +
    c_2\left(\lambda-\lambda_0\right) + 1\right].
\end{equation}

Here, $\lambda_0$ = 350~\micron, and for both models, $a$ is an overall
normalization constant (it is the \emph{fitted} value of $p_{350}$). The models
are intended to probe the shape of the per-pixel polarization spectra over
the wavelength range of 250~\micron to 850~\micron. The power-law
model investigates whether the spectra are increasing or decreasing. The 
quadratic model, in addition, allows for the polarization spectra to have
minima or maxima somewhere within this wavelength range. 
\begin{figure}[htbp!]
%\begin{figure}[p]
\epsscale{1.0}
\plotone{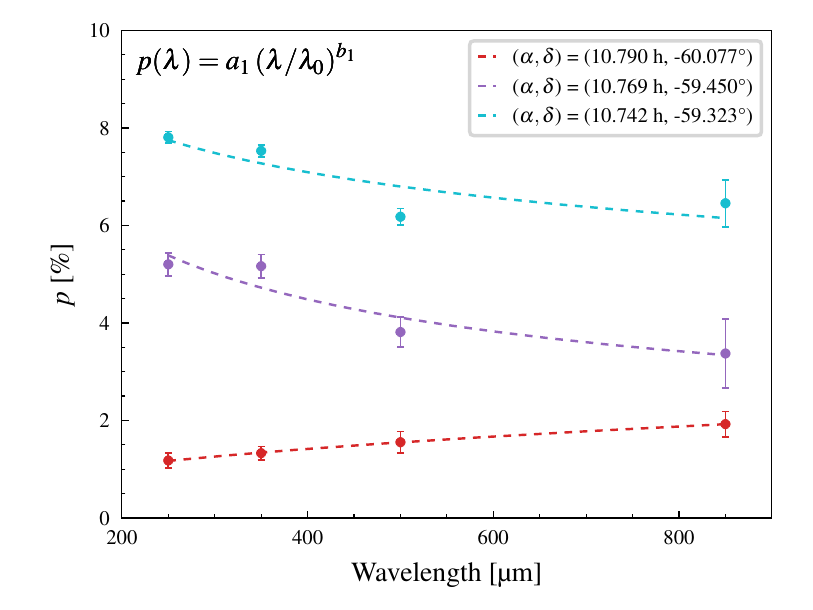}
\plotone{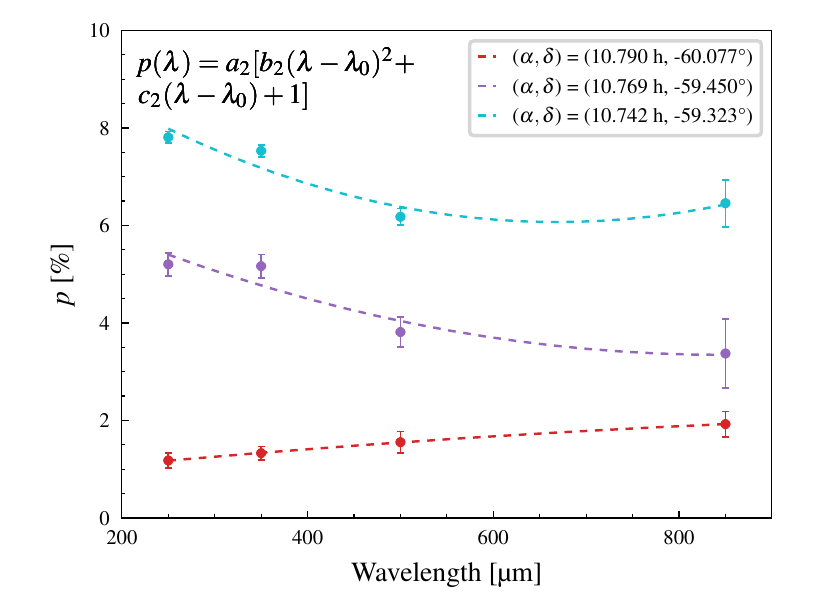}
    \caption{Power-law (top) and quadratic (bottom) fits to $p(\lambda)$
    for three example pixels. The plots are shown for diffuse
    emission subtraction using the Far reference region.\label{fig:eg_pix}}
\end{figure}

\begin{table*}[ht!]
\begin{center}
    \caption{Medians and MADs of $p(\lambda)$ Fit
    Parameters\label{tab:fit_params}}
\begin{tabular}{lccccc}
\hline
\hline 
    Diffuse Emission&\multicolumn{2}{c}{Power Law
    Fit}&\multicolumn{3}{c}{Polynomial Fit}\\
    Subtraction Method&$b_1$ &$p_{350}/a_1$&$b_2$
    ($\times 10^{-6}$)&$c_2$ ($\times 10^{-4}$)&$p_{350}/a_2$ \\
\hline
    Far&$-0.01 \pm 0.10$&$1.01 \pm 0.03$&$0.8 \pm 1.3$&$-3.1 \pm 4.9$ &$1.03 \pm 0.03$ \\
    Near&$-0.04 \pm 0.10$&$1.01 \pm 0.03$&$0.9 \pm 1.4$&$-4.2 \pm 4.6$&$1.03 \pm 0.03$\\
    None&$-0.14\pm 0.10$&$0.96 \pm 0.04$&$1.1 \pm 1.1$&$-8.0 \pm 4.6$&$0.96 \pm 0.03$\\
\hline
\end{tabular}
\end{center}
\end{table*}

\begin{figure}[htb!]
%\begin{figure}[p]
\epsscale{1.0}
\plotone{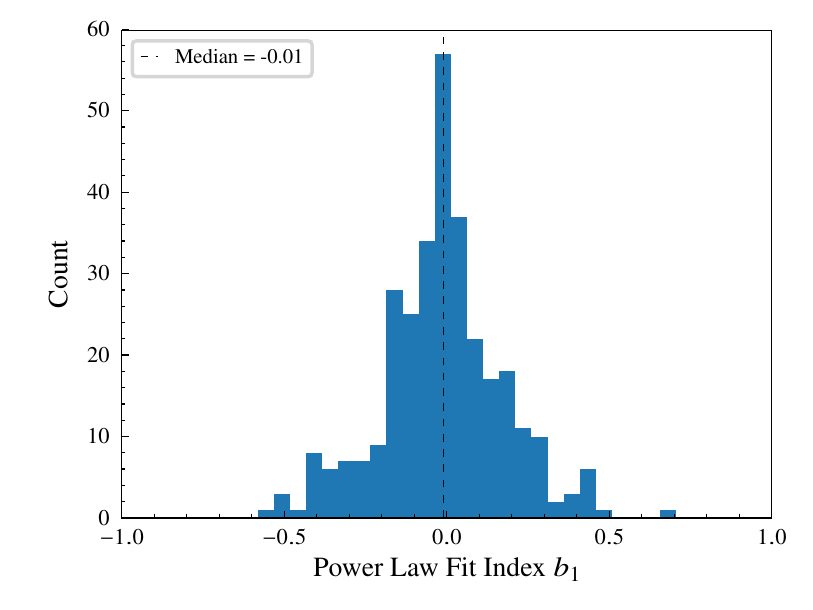}
    \caption{Histogram of the power-law fit parameter $b_1$ from
    Equation~\ref{eq:pwrlaw}.\label{fig:pwr_dist}}
\end{figure}

\begin{figure}[htb!]
\epsscale{1.0}
\plotone{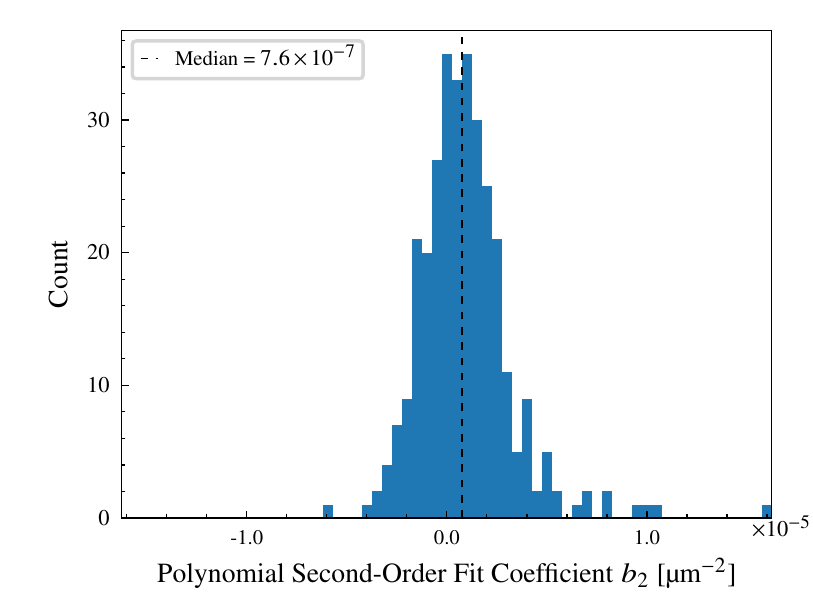}
\plotone{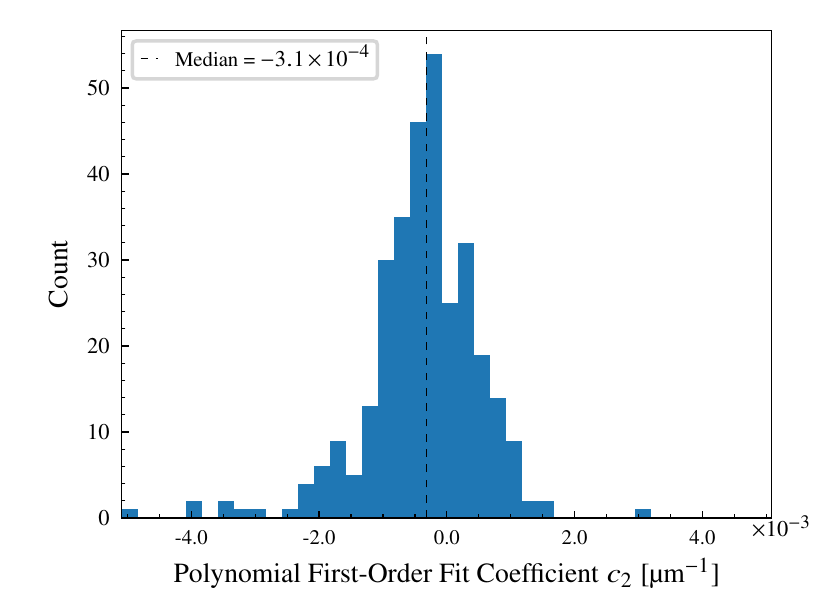}
    \caption{Histograms of the polynomial fit parameters $b_2$ (top) and
    $c_2$ (bottom) from
    Equation~\ref{eq:quadlaw}.\label{fig:quad_dist}}
\end{figure}

Figure~\ref{fig:eg_pix} shows the results of the power-law and quadratic
fitting for three example pixels. In these plots, the error bars in each
band are derived from the TOAST covariances (Appendix~\ref{sec:err}).
After fitting to the per-pixel spectra, the distributions of the fit parameters were analyzed. The medians 
and MADs are listed in Table~\ref{tab:fit_params} for the fit parameters
relevant to the spectral shape, i.e.~$b_1$, $b_2$, and $c_2$. This table also
lists the median $\pm$ MAD values of the ratios $p_{350}/a$, which indicate the extent to which
the model value for the fractional polarization at 350~\micron matches the
measured value. The distributions of the spectral-shape fit parameters for
the Far reference region case are
shown in Figure~\ref{fig:pwr_dist} for the power-law fit, and in
Figure~\ref{fig:quad_dist} for the quadratic fit. 

\begin{figure*}[htbp!]
%\begin{figure*}[p]
\epsscale{1.0}
%\plotone{figure1.pdf}
\plotone{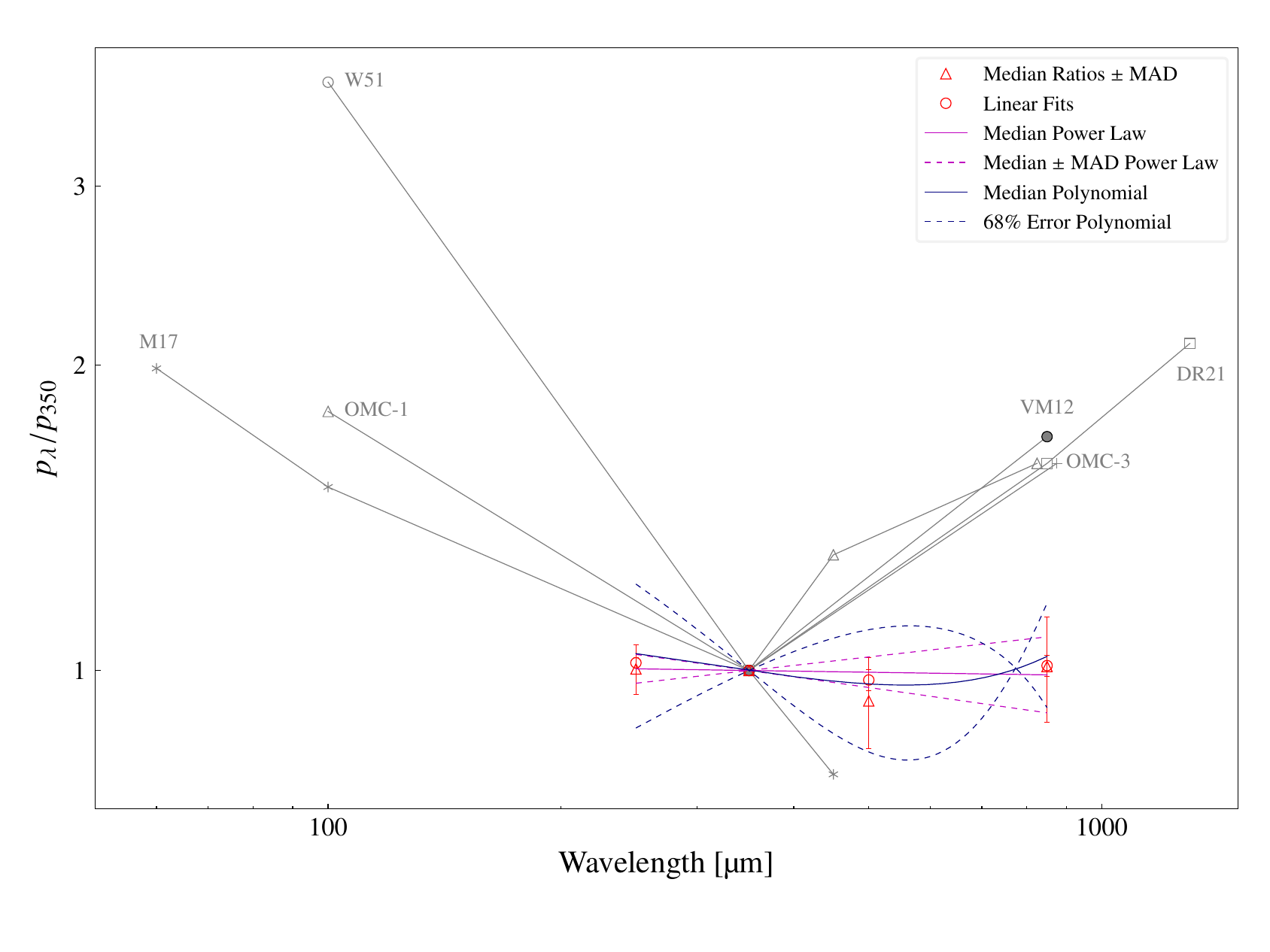}
    \caption{Polarization spectra from previous work in gray, with new
    BLASTPol/\Planck Carina data added in color. The gray data points at
    850~\micron have been offset horizontally to distinguish them. The W51
    and OMC-1 $p_{100}/p_{350}$ values and the DR21 $p_{1300}/p_{350}$ 
    values are from~\citet{vaill02}. All
    previous measurements at 850~\micron are from~\citet{vaill12}. The solid circle (VM12)
    is the median of these 850~\micron measurements over 15 clouds. The OMC-1
    $p_{450}/p_{350}$ value is from~\citet{vaill08}. The M17 points are from~\citet{zeng13}. 
    The red triangles
    show the median polarization ratios with MAD error bars. The red
    circles are the best-fit slopes of linear fits to scatter plots of
    $p_\lambda$ versus $p_{350}$. The magenta lines show spectra produced using the
    power-law fit parameters, while the dark blue lines show spectra
    produced using the quadratic fit parameters. For these two cases, the
    solid lines are spectra produced using the median fit parameters,
    while the dashed lines reflect the distribution in the fit parameters
    (see text). \label{fig:money}}
\end{figure*}

\subsection{Summary of Polarization Spectrum Measurements}
    
The result of a flat spectrum for Carina computed using all the
    previously-described methods is shown in Figure~\ref{fig:money} for
    the case of diffuse emission subtraction using the Far reference
    region. The median polarization ratios $\pm$ MADs
    (Section~\ref{sec:polratios}) are shown as red triangles. The polarization
    ratios from linear fits to scatter plots of $p_\lambda$ versus
    $p_{350}$  are shown as red circles, with error bars based on
    bootstrap resampling (Section~\ref{sec:scatter}). Representative
    power-law and quadratic fits to per-pixel polarization spectra
    (Section~\ref{sec:model}) are also shown.
    For the power-law fit, the mean and dispersion among the per-pixel polarization
spectra are demonstrated by plotting the power-law model corresponding to 
    $\mathrm{median}(b_1)$ as a solid magenta line, and plotting the
    models corresponding to $\mathrm{median}(b_1)
    \pm \mathrm{MAD}(b_1)$ as dashed magenta lines. Similarly, for the
    second-order polynomial fit, the parabola corresponding to the median
    values of $b_2$ and $c_2$ is plotted as a solid dark blue line.
However, the fit parameters $b_2$ and
$c_2$ are highly anti-correlated, with Pearson correlation coefficient
$\rho_{c_2,b_2} = -0.92$ for the Far reference region case. Therefore, it
was not sufficient to simply plot extremal models using the median $\pm$
MAD values of each fit parameter individually. The 68\% error ellipse of
their joint distribution was constructed by diagonalizing the covariance
matrix of $b_2$ and $c_2$ using eigenvalue decomposition. The major axis
of the resulting ellipse had endpoints at $(c_2, b_2) = (-1.8 \times
10^{-3}, 4.2 \times 10^{-6})$ and at $(c_2, b_2) = (9.9 \times 10^{-4},
    -2.3 \times 10^{-6})$. The dashed dark blue lines in
    Figure~\ref{fig:money} correspond to the parabolas having these extremal
    fit parameter values.
    
\subsection{Effect of Environment}
\label{sec:env}
\begin{figure*}[ht!]
\epsscale{1.0}
\plotone{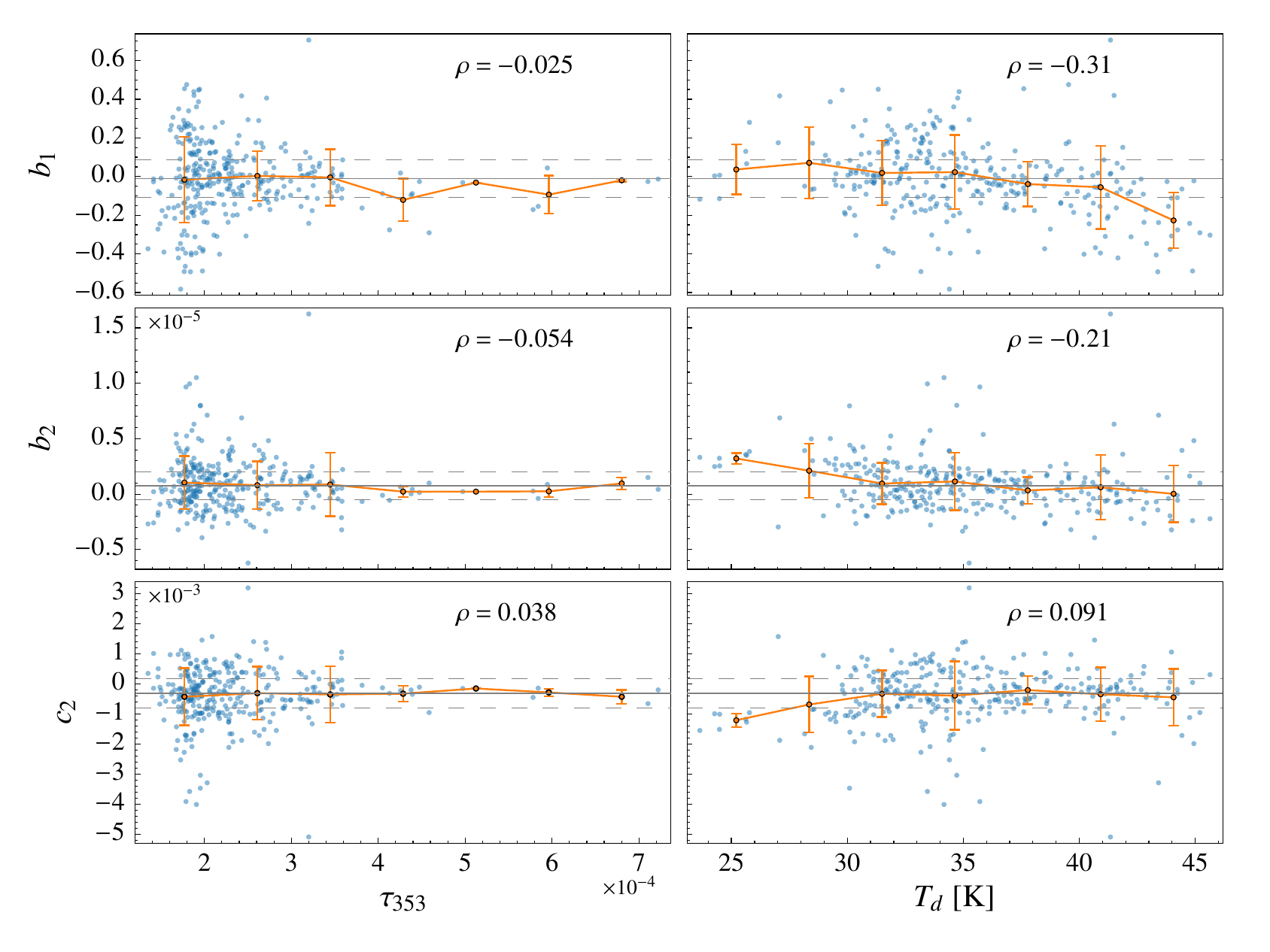}
    \caption{Plots of polarization spectrum shape parameters versus dust
    optical depth $\tau_{353}$ (left column) or versus dust temperature $T_d$
    (right column) for the case of diffuse emission subtraction using the
    Far reference region. From top row to bottom, the parameters are,
    respectively, the spectral index $b_1$ of the power-law fit, the
    second-order coefficient $b_2$ of the quadratic fit, and the
    first-order coefficient $c_2$ of the quadratic fit. The light blue
    points show the data for every sightline, while the orange points and error bars show the mean and standard deviation of each shape parameter within bins linearly-spaced in $\tau_{353}$ or $T_d$. The horizontal gray lines show the median (solid) and median $\pm$ MAD (dashed) parameter values over the whole cloud.\label{fig:env}}
\end{figure*}

An investigation was undertaken to probe whether the shape of the
    polarization spectrum over these four bands exhibits any dependence on
    the molecular cloud environment. Polarization spectrum parameters were
    correlated with two environmental parameters: dust temperature $T_d$,
    and dust optical depth at 353~GHz $\tau_{353}$. The latter is
    proportional to the dust column density. These parameters were obtained from the \Planck all-sky dust model \citep{planck_xi14} first mentioned in Section~\ref{sec:cal}. This model is generated by fitting a modified blackbody SED to the high-frequency dust $I$ maps from HFI at 353, 545, and 857~GHz, along with a highest-frequency map from IRAS 100~\micron data. This SED is of the form
\begin{equation}
\label{eq:mbb}
I_\nu(\nu) = \tau_{\nu_0}\left(\frac{\nu}{\nu_0}\right)^{\beta_d}B_\nu(\nu, T_d).
\end{equation}
In Equation~\ref{eq:mbb}, $\beta_d$ is the power-law spectral index of the frequency-dependent dust emissivity, $B_\nu$ is the Planck function, and $T_d$ is the dust temperature. The parameter $\tau_{\nu_0}$ is the dust optical depth at a reference frequency of $\nu_0 = 353~\si{\giga\hertz}$. In~\citet{planck_xi14}, it is emphasized that these three parameters are only approximations to the true dust properties. A single-component model has been assumed, whereas in reality multiple temperature components could exist along any given line of sight. Therefore the model parameters $\tau_{353}$ and $T_d$ are used here only to establish the relative ordering among sightlines in order to search for very obvious trends, which, if present, would lend themselves to a more detailed future investigation.  
\citet{gand16} used a similar dust SED model fitted to the \Herschel-SPIRE
160, 250, 350, and 500~\micron maps of Vela C. This fit assumed $\beta_d =
2$ and obtained values for $T_d$ and column density $N_H$. The per-pixel
polarization spectrum shape parameters $b_1$, $b_2$, and $c_2$ defined in
Section~\ref{sec:model} were plotted against $T_d$ and $N_H$ in order to
search for a dependence of spectral shape on environment. For the Carina
Nebula however, due to the limited spatial extent of the available
\Herschel maps, it was not possible to carry out such a fit using the
exact same reference region for diffuse emission subtraction as was used
for the BLASTPol maps. Thus, we opted to use the \Planck all-sky dust
model instead. Using the \Planck-derived $\tau_{353}$ as the environmental
density parameter rather than computing $N_H$ also avoids making
assumptions about the dust opacity, $\sigma_{353} = \tau_{353}/N_H$,
within our region. 

Figure~\ref{fig:env} shows the results of plotting the power-law and
    polynomial fit parameters versus temperature and optical depth. In
    addition to the points for every individual sightline, the fit
    parameters are binned into seven evenly-spaced bins in $T_d$ or
    $\tau_{353}$. The binned curves appear quite flat. For the most part,
    the mean values of the fit parameters within each bin (orange points)
    lie well within the median $\pm$ MAD range of the fit parameters over
    the whole cloud (bounded by the gray dashed lines). Exceptions to this are the first $T_d$ bin for $c_2$ and $b_2$, and the last $T_d$ bin for $b_1$. However, these bins only contain a handful of points each. Each panel of Figure~\ref{fig:env} also shows the value of the Pearson correlation coefficient $\rho$ for the scatter plot in question. Overall, strong evidence for trends in polarization spectrum shape with environment are not observed.

\section{Discussion}
\label{sec:disc}

As shown in Figure~\ref{fig:money}, the result of a polarization spectrum
    that is flat to within $\pm$15\% is inconsistent with the results of
    measurements of a number of molecular clouds obtained by ground-based
    submillimeter telescopes. Recall (Section~\ref{sec:intro}) that the
    BLASTPol/\Planck submillimeter polarization spectrum for the Vela C
    GMC presented by~\citet{gand16} was also much flatter than the
    ground-based spectra.  As a potential explanation for this discrepancy,
    \citet{gand16} posited that ground-based observations were limited to
    the densest sightlines within molecular clouds, in a very different
    regime of column density compared to the cloud-averaged case for
    BLASTPol. This can be quantified using the \Planck 850~\micron
    intensity as a proxy for column density. Although this proxy (unlike
    $\tau$) is not completely free from dependence on dust temperature, it
    was the only method available for comparing column density between the
    BLASTPol and ground-based measurements. Table~\ref{tab:cold} below
    shows a comparison of the median and interquartile range of
    the 850~\micron intensity computed for Carina (this work), for Vela
    C~\citep{gand16}, for a translucent molecular cloud~\citep{ashton18},
    and for ground-based measurements of 17 molecular
    clouds, which were calculated by~\citet{gand16} using online data provided
    by~\citet{vaill12}.
\begin{table*}[htbp!]
    \caption{Comparison of quartiles in 850~\micron intensity in MJy\,sr$^{-1}$ between \Planck measurements and ground-based measurements of molecular clouds. The Carina Nebula values are computed over the sightlines passing the data cuts for the case of diffuse emission subtraction using the Far reference region.
\label{tab:cold}}
\begin{center}
\begin{tabular}{lccc}
\hline
\hline 
& $1^\mathrm{st}$ Quartile & Median & $3^\mathrm{rd}$ Quartile\\
\hline 
Translucent Cloud~\citep{ashton18} & 3.1 & 3.4 & 4.1\\ 
Vela C~\citep{gand16} & 6.5 & 9.1 & 14.1\\ 
Carina Nebula~(this work) & 7.6 & 10.8 & 17.6\\ 
Ground-based, 17 molecular clouds~\citep{vaill12} & 300 & 637 & 1327\\ 
\hline
\end{tabular}
\end{center}
\end{table*}
The comparison shows that the median 850~\micron intensity for the Carina Nebula is quite similar to that of the Vela C measurement, indicating a broadly similar regime in column density. In comparison, the median intensity of the ground-based measurements is nearly two orders of magnitude higher.

Another interesting comparison is to the analysis of the most recent
\Planck data release (PR3) in ~\citet{planck_xi_18}. The equivalent information
to the polarization spectrum is reported by \Planck in terms of the difference in
power law spectral indicies of the SEDs for total intensity and for
polarization, $\beta_d^I$ and $\beta_d^P$. Under the simplistic assumption
that the dust grains contributing to polarized and to overall emission are
all isothermal at temperature $T_d$, these two SEDs can be written as
$I(\nu) \propto \left(\nu/\nu_0\right)^{\beta_d^I}B_\nu(\nu, T_d)$ and
$P(\nu) \propto \left(\nu/\nu_0\right)^{\beta_d^P}B_\nu(\nu, T_d)$. It can
then be shown that the quantity computed in this work, the polarization
ratio, is given by
\begin{equation}
    \frac{p(\nu)}{p(\nu_0)} = \left(\frac{\nu}{\nu_0}\right)^{\beta_d^P -
    \beta_d^I}.
\end{equation}
Therefore, a measurement of $\beta_d^P < \beta_d^I$ would imply a
polarization spectrum that was falling with frequency, or rising with
wavelength. In the analysis of the PR3 maps, polarized emission was 
investigated for six nested sky patches at high Galactic latitude. A model
was fitted to the angular cross power spectra between \Planck frequency
maps. This model included a power-law dependence on multipole moment ($\ell$),
a power-law frequency dependence for polarized synchrotron emission, and a
modified blackbody SED with power-law emissivity for polarized dust
emission at an assumed dust temperature $T_d = 19.6~\si{\kelvin}$. 
The result, averaged over all sky patches and
multipole bins, was $\beta_d^P = 1.53 \pm 0.02$. No
statistically-significant difference from the spectral index for total
intensity was found: $\beta_d^P - \beta_d^I = 0.05 \pm
0.03$~\citep{planck_xi_18}. This result
implies a flat polarization spectrum, on average, at millimeter wavelengths.

\begin{figure}[htbp!]
\epsscale{1.0}
\plotone{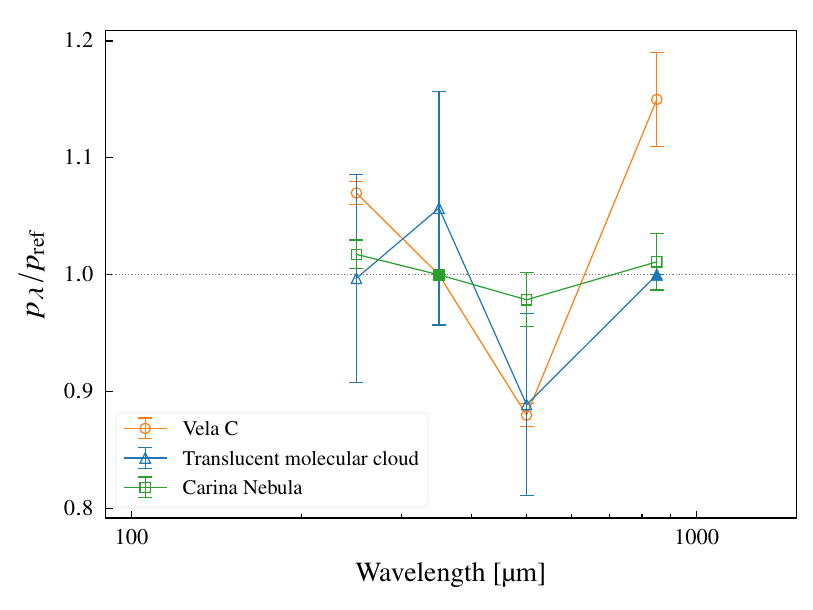}
    \caption{Examples of three polarization spectra from the BLASTPol 2012
    flight. Shown are measurements of a translucent molecular cloud near
    to Vela C on the sky~\citep[][triangular points]{ashton18}, measurements
    of the Vela C molecular cloud~\citep[][circular points]{gand16},
    and measurements of the Carina Nebula (this work, square points). For
    Vela C and Carina, the
    points shown, and the associated error bars, are from linear fits to scatter
    plots of $p_\lambda$ versus $p_\mathrm{ref}$, with a reference band of
    350~\micron. For the translucent cloud, the points shown, and the
    associated statistical error
    bars, come from linear fits to $Q_\lambda$ and $U_\lambda$ \emph{measured} in the
    BLASTPol bands versus $Q_\lambda$ and $U_\lambda$ 
    \emph{extrapolated} to those bands (from a reference
    band of 850~\micron) using the \Planck all-sky dust model.
    The filled points show which band is the reference band for each
    polarization spectrum plotted.  
    \label{fig:bpcompare}}
\end{figure}

In addition to this overall agreement with the high-latitude diffuse
emission, there is now striking agreement among BLASTPol molecular cloud
targets in the Galactic plane. The result of a flat spectrum in polarization fraction has now been reproduced for measurements of a cold, dense molecular cloud~\citep{gand16}, a translucent molecular cloud~\citep{ashton18} and a more evolved, warmer molecular cloud being actively-heated by many internal stellar sources (this work). A direct comparison of these three results is shown in Figure~\ref{fig:bpcompare}. Some care should be taken in the interpretation of this figure, since the polarization spectrum from~\citet{ashton18} is normalized to the 850~\micron band, whereas the other two polarization spectra are normalized to 350~\micron. Even so, despite the differing radiative environments and densities of these targets, the result of a flat submillimeter polarization spectrum persists as a common property of these molecular clouds.

\begin{figure}[htbp!]
\epsscale{1.0}
\plotone{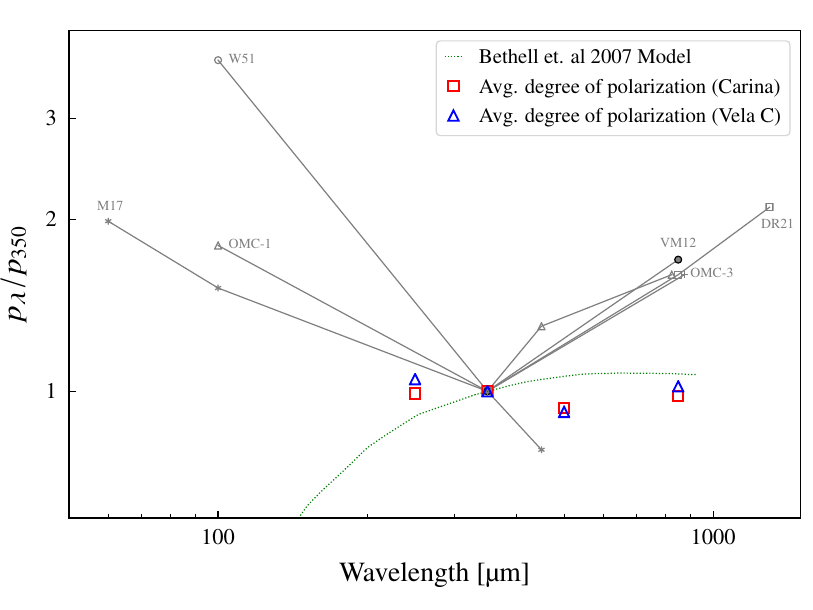}
    \caption{Comparison with the polarization spectrum predicted
    by~\citet{bethell07}, shown as a green dotted line. The red squares
    represent the total polarized fraction of the Carina data (see text)
    normalized to 350~\micron, while the blue triangles show the same quantity for Vela C, as reported in~\citet{gand16}. \label{fig:bethell}}
\end{figure}

This result of a lack of significant wavelength-variation in the
polarization spectrum for a cloud such as the Carina GMC, with significant internal heating from
embedded sources, potentially has implications for theoretical grain
alignment mechanisms. For example, the RAT mechanism
(Section~\ref{sec:intro}) predicts a higher grain alignment efficiency for
dust grains that are less shielded from stellar radiation~\citep{andersson15}. However, a detailed investigation of the
implications of our result to grain-alignment theory is beyond the scope
of this work. For the purpose of a baseline comparison to theory, we
examine the model of a cold, dense molecular cloud with no internal
sources presented in~\citet{bethell07}. Figure~\ref{fig:bethell} shows
this model over the wavelength range of interest. Overlaid are the
previous ground-based polarization spectrum measurements first shown in
Figure~\ref{fig:money}. In order to match the analysis method
of~\citet{bethell07}, the
BLASTPol/\Planck data for Carina, and from Vela C from~\citet{gand16}, are shown here as the ``total polarized fraction'', which is computed as $\left. \left(\sum_j P_j\right) \middle/ \left(\sum_j I_j\right)\right.$, where $j$ indexes the pixels passing the data cuts. The results for these two BLASTPol targets are broadly consistent with the Bethell model, certainly showing a much closer correspondence than than the previous ground-based studies.

The negative slope of the ground-based polarization spectra in the
far-infrared has been explained heuristically in terms of a very strong
ETAC effect (Section~\ref{sec:intro}), perhaps due to internal sources, 
in which hotter, less-shielded dust
grains are preferentially more aligned, leading to a higher polarization
fraction for the dust emission at shorter wavelengths~\citep{hildebrand99, zeng13}.
The righthand side of the V is more problematic, however: slopes of this 
steepness are not reproduced in the BLASTPol data. Furthermore, no
theoretical models have been constructed that can reproduce the very large
polarization ratios that are seen, in general, in the ground-based
spectra. Regardless, we can conclude that the cloud-averaged submillimeter
polarization spectra of two very different molecular clouds agree with
the flat~\citet{bethell07} prediction for externally-illuminated, dense
molecular clouds with no internal radiation sources. Thus, it appears, at
least from the BLASTPol measurements, that
the internal sources do not significantly affect the \emph{cloud-averaged}
submillimeter polarization spectrum.

In light of the discrepancy between BLASTPol and ground-based
submillimeter polarization spectrum results, a logical step for future work
would be to repeat the measurements of the specific ground-based targets
using more advanced ground-based polarimeters of higher sensitivity
and mapping speed that are slated to come online in the near future. One
example is the TolTEC camera~\citep{bryan18}, which will probe wavelengths
of 1~mm and longer. Another interesting avenue for future research is a
more detailed investigation of the spatial variation of the polarization
spectrum from point to point within a cloud, and the correlation of the
spectral shape with environment. The BLASTPol data show some evidence of
spatial variation (Figure~\ref{fig:pol_ratio_maps}), but no strong trends
with environment. Ultimately this investigation is limited by the angular
resolution of the data. The BLAST-TNG experiment~\citep{tyr14b}, which is
currently scheduled for an Antarctic balloon flight in the Austral summer
of 2018-2019, will greatly aid this effort. This polarimeter offers an
order of magnitude more detectors than BLASTPol and will observe in the
same bands with a much higher angular resolution of 31$^{\prime \prime}$
to 59$^{\prime\prime}$ FWHM. The combination of high resolution and full cloud-scale coverage offered by this experiment may also help test the hypothesis stated above that the discrepancy between the BLASTPol/\Planck and ground-based measurements is due to the fact that the latter were only sensitive to the densest clumps within GMCs.

\section{Summary}
\label{sec:summ}

Measurements of the linear polarization along 314
sightlines were made by BLASTPol in the
Carina Nebula in the 250, 350, and 500~\micron wavebands. These data were
combined with \Planck 353 GHz (850~\micron) data from the same region
in order to produce submillimeter polarization spectra. These spectra were 
calculated using several methods, including
the median polarization ratios, slopes from linear fits to scatter plots
of $p_\lambda$ versus $p_{350}$, and by fitting quadratic and power-law
models $p(\lambda)$ to the per-pixel polarization spectra. No strong
evidence was found for variation of the fitted parameters of these models
as a function of cloud environment, as quantified by the dust temperature
$T_d$ and dust optical depth $\tau_{353}$, which were obtained from the
\Planck all-sky dust model. The cloud-averaged 
polarization spectrum of the Carina Nebula
appears flat to within $\pm$15\% in the polarization ratio quantity
$p_\lambda / p_{350}$, where $p_\lambda$ is the fractional linear
polarization in a given waveband. This is at odds with previous
ground-based measurements of the polarization spectrum of other molecular
clouds, which showed a V-shaped spectrum, with a negative slope in the
far-infrared, a positive slope towards millimeter wavelengths, and a
pronounced minimum near 350~\micron. The flatness of the polarization
spectrum in Carina is, however, in remarkably close agreement with
BLASTPol/\Planck measurements in other molecular clouds, including the
measurement of the Vela C GMC~\citep{gand16}, and of a translucent
molecular cloud near to Vela C on the sky~\citep{ashton18}. The shapes of
the Vela C and Carina polarization spectra are both in relatively good 
agreement with the~\citet{bethell07} theoretical prediction for an
externally-illuminated, dense molecular cloud with no internal radiation sources.

\acknowledgments{The BLASTPol collaboration acknowledges support from
  NASA (through grant numbers NAG5-12785, NAG5-13301,
  NNGO-6GI11G, NNX0-9AB98G, and the Illinois Space Grant
  Consortium), the Canadian Space Agency, the Leverhulme
  Trust through the Research Project Grant F/00 407/BN,
  Canada's Natural Sciences and Engineering Research Council,
  the Canada Foundation for Innovation, the Ontario Innovation
  Trust, and the US National Science Foundation Office of Polar
  Programs. This work was based in part on observations obtained with \Planck (http://www.esa.int/Planck), an ESA science mission with instruments and contributions directly funded by ESA Member
  States, NASA, and Canada. C.B.N.~also acknowledges support
  from the Canadian Institute for Advanced Research. F.P.S.~is
  supported by the CAPES grant 2397/13-7. F.P.~acknowledges the European 
  Union's Horizon 2020 research and innovation programme under grant
  agreement number 687312 (RADIOFOREGROUNDS). P.A.~is
  supported through Reach for the Stars, a GK-12 program
  supported by the National Science Foundation under grant
  DGE-0948017. L.M.F.~is a Jansky Fellow of the National Radio Astronomy 
  Observatory (NRAO). NRAO is a facility of the National Science Foundation 
  (NSF), operated under cooperative agreement by Associated Universities, Inc. Finally, we thank the Columbia Scientific Balloon Facility staff for their outstanding work.
}

\appendix
\section{Correction of a Large-Scale Systematic Using Herschel Intensity
Maps}
\label{sec:herschel}

During the initial polarization spectrum analysis of Carina, anomalously-high
values were found for the polarization ratio $p_{850}/p_{350}$ for sightlines
in the northern part of the cloud. The problem was determined to be in the BLASTPol data:
receiver 1/$f$ noise led to the BLASTPol $I$ maps having a large-scale gradient in the
north-south direction, which is perpendicular to the predominant scan direction during Carina observations. Systematically-high values of $I_{250}$, $I_{350}$, and $I_{500}$
led to systematically-low values of $p_{250}$, $p_{350}$, and $p_{500}$ in the 
north of the cloud. This bias in turn inflated the \Planck-to-BLASTPol polarization ratio. Since the striping is predominantly in $I$, it could be corrected by differencing calibrated BLASTPol maps with publicly-available \Herschel-SPIRE maps taken of Carina in the same bands. 

The \Herschel maps were first re-gridded to match the BLASTPol
pixelization. A least-squares fit of a fifth-order polynomial was applied
to the  (BLASTPol $-$ \Herschel) intensity values (for pixels within the
left Far reference region rectangle) as a function of pixel
$y$-coordinate. The fitted intensity as a function of map $y$-coordinate was extended horizontally across the entire map, to produce a model for the vertical gradient. This model was then subtracted from the BLASTPol $I$ maps in each band. 

Only pixels within the left Far reference region rectangle were used in
the fit, because there was a slight beam mismatch between the smoothed
BLASTPol and smoothed \Herschel intensity maps, leading to leakage of
cloud structure (including intensity peaks) into the (BLASTPol $-$
\Herschel) difference map. The left Far reference region rectangle happens
to be free of such structure, meaning that most of the flux variation
within it is due to the gradient alone. As the analysis progressed, more
refined data cuts excluded some of the more extreme northern sightlines
from the final data set. As a result, repeating the analysis of this paper
with diffuse emission subtraction using the Far refrence region, and
\emph{without} the \Herschel correction, results in relatively small
changes to the median ratios of $\Delta\left(p_{\lambda}/p_{350}\right)$ =
$-$0.007, +0.019, and $-$0.048 at 250, 500, and 850~\micron, respectively. Since the systematic does not change the final result of a polarization spectrum
that is flat to within $\pm 15\%$ over all four bands, it was not deemed to be necessary to repeat the \Herschel correction more carefully using maps of matching resolution.

\section{Error Propagation}
\label{sec:err}

The per-pixel variances of the polarization quantities, $P$, $p$, and
$\psi$  are computed from the
per-pixel covariance matrix of the Stokes parameters on the sky (as
estimated by the TOAST map maker) using the following procedure. Each of
the above three quantities can be expressed as a function $f(I,Q,U)$. If
this function is approximated by its first-order Taylor series expansion
about the mean of the pixel Stokes parameter values, then the variance in
$f$ is given by:

\def\arraystretch{2}
\begin{equation}
    \label{eq:covar}
    \sigma^2_f = \left[ \begin{array}{ccc} \frac{\partial f}{\partial I} & \frac{\partial f}{\partial Q} & \frac{\partial f}{\partial U} \end{array}\right] \left[\begin{array}{ccc} \sigma_I^2 & \sigma_{IQ} & \sigma_{IU} \\ \sigma_{IQ} & \sigma_{Q}^2 & \sigma_{QU} \\ \sigma_{IU} & \sigma_{QU} & \sigma_{U}^2 \end{array}\right]\left[ \begin{array}{c} \frac{\partial f}{\partial I} \\ \frac{\partial f}{\partial Q} \\ \frac{\partial f}{\partial U} \end{array}\right].
\end{equation}

Applying Equation~\ref{eq:covar} to each of Equations~\ref{eq:P}, \ref{eq:pfrac}, and \ref{eq:psi} above yields the following results for the variances of the polarization quantities:

\begin{equation}
    \sigma_P^2 = \frac{1}{P^2}\left[\left(Q^2\right) \sigma_Q^2 + \left(U^2\right) \sigma_U^2 + \left(2QU\right) \sigma_{QU}\right],
\end{equation}

\noindent and

\begin{equation}
    \sigma_{\psi}^2 = \left(\frac{180\si{\degree}}{\pi}\right)^2 \frac{1}{4P^4}\left[\left(U^2\right) \sigma_Q^2 + \left(Q^2\right) \sigma_U^2 - \left(2QU\right) \sigma_{QU}\right].
\end{equation}
For notational convenience, we can define \emph{normalized} Stokes parameters for linear polarization, given by $q \equiv Q/I$ and $u \equiv U/I$. Applying Equation~\ref{eq:covar} to these, we have
\begin{equation}
    \sigma_q^2 = q^2 \left( \frac{\sigma_I^2}{I^2} + \frac{\sigma_Q^2}{Q^2} - 2 \frac{\sigma_{IQ}}{IQ}\right),
\end{equation}

\begin{equation}
    \sigma_u^2 = u^2 \left( \frac{\sigma_I^2}{I^2} + \frac{\sigma_U^2}{U^2} - 2 \frac{\sigma_{IU}}{IU}\right),
\end{equation}

\noindent and

\begin{equation}
    \sigma_{qu} = qu \left( \frac{\sigma_I^2}{I^2} + \frac{\sigma_{QU}}{{QU}} - \frac{\sigma_{IQ}}{IQ} - \frac{\sigma_{IU}}{IU} \right).
\end{equation}
The variance in $p$ can therefore be expressed compactly as
\begin{equation}
    \sigma_p^2 = \frac{1}{p^2}\left[\left(q^2\right)\sigma_q^2 + \left(u^2\right)\sigma_u^2 + \left(2qu\right)\sigma_{qu}\right].
\end{equation}

\end{document}